\documentclass[a4paper,11pt]{article}%
\usepackage{amssymb}
\usepackage{graphicx}
\usepackage{amsbsy}
\usepackage{amsmath}
\usepackage{cite}
\usepackage{slashed}
\usepackage{amsfonts}
\usepackage{hyperref}%
\hypersetup{
colorlinks   = true,
linkcolor    = blue,
citecolor    = blue,
}
\newcommand*{\doi}
[1]{\href{http://dx.doi.org/#1}{doi: #1}}
\begin{document}

\thispagestyle{empty} \setcounter{page}{0}
\begin{flushright} December 2024\\
\end{flushright}

\vskip3.4 true cm

\begin{center}
{\huge Dark-matter induced neutron-antineutron oscillations}\\[1.9cm]

\textsc{Theo Brugeat}$^{1}$\textsc{, Christopher Smith}$^{2}$\vspace
{0.5cm}\\[9pt]\smallskip{\small \textsl{\textit{Laboratoire de Physique
Subatomique et de Cosmologie, }}}\linebreak%
{\small \textsl{\textit{Universit\'{e} Grenoble-Alpes, CNRS/IN2P3, Grenoble
INP, 38000 Grenoble, France}.}} \\[1.9cm]\textbf{Abstract}\smallskip
\end{center}

\begin{quote}
\noindent If dark matter carries a baryon number of two, neutron-antineutron
oscillations could require its presence to manifest themselves. If it is in
addition very light, in the micro-eV range or up to a few orders of magnitude
below, these oscillations could even exhibit a Rabi resonance. Though the
magnetic tuning required to convert a macroscopic number of neutrons into
antineutrons is not realistic, sizeable enhancements remain possible. Building
on this observation, axionic realizations for this scenario are systematically
analyzed. For true QCD axion models, we find that the Goldstone boson nature
of the axion imposes the presence of axionless $n-\bar{n}$ mixing effects,
either in vacuum or in decays, which are sufficiently constrained
experimentally to leave no room for axion-induced oscillations. Thus, a
generic scalar or axion-like dark matter background would have to exist to
induce resonant $n-\bar{n}$ oscillations. Yet, if Nature has taken that path
to relate dark matter and baryon number violation, the experimental signature
would be striking and certainly worth pursuing.

\let\thefootnote\relax\footnotetext{\newline$^{1}\;$%
brugeat@lpsc.in2p3.fr\newline$^{2}$~chsmith@lpsc.in2p3.fr}
\end{quote}

\newpage

\setcounter{tocdepth}{2}
\tableofcontents

\section{Introduction}

Among cosmological enigmas, the nature of dark matter and the origin of the baryonic asymmetry of the Universe stand out in the particle physics community. Answering them should be possible without departing too much from our current understanding of the microscopic world. After all, baryon number $\mathcal{B}$ is only accidentally conserved in the Standard Model Lagrangian, and does not survive to quantization. At the same time, the extraordinary stability of the proton seems at odd with only an approximate conservation of baryon number. One way out here is that proton decay requires both baryon and lepton number $\mathcal{L}$ to be broken by one unit. By contrast, the smallness of neutrino masses rather points towards a violation of $\mathcal{L}$ by two units. It is then quite natural to imagine those are accompanied by $\Delta\mathcal{B}=2$ interactions, given that $\mathcal{B}-\mathcal{L}$ is anomaly free and could even be gauged in a grand unified context. This will be our first starting point. The second one is the observation that the dark matter and baryon relic densities could be magnitudes apart, but happen to be of the same order. Though we will not in any way attempt to construct a model explaining this possibly coincidental fact, we will use it as a motivation to entangle the two cosmological puzzles. In practice, our strategy to do that is to ask dark matter to have a baryon number of two, and then study the possible signatures of such a scenario.

At low-energy, neutron-antineutron oscillations are the hallmark of $\Delta\mathcal{B}=2$ interactions~\cite{Mohapatra:1980de}. Let us recall their main features (see e.g. Refs.~\cite{Phillips:2014fgb,Mohapatra:2009wp} for reviews). Typically, some UV physics at the scale $\Lambda_{\Delta\mathcal{B}=2}$ induces six-quark effective interactions, whose hadronization collapses to Majorana mass terms for the neutron%
\begin{equation}
\mathcal{L}_{free}=\bar{n}(i \!\not\!
\partial-m)n-\frac{1}{2}\varepsilon(\bar{n}^{\mathrm{C}}n+\bar{n}%
n^{\mathrm{C}})\ .\label{Intro1}%
\end{equation}
Given their dimension-nine origin, we expect the mixing term to be tiny,
$\varepsilon/m\sim(\Lambda_{QCD}/\Lambda_{\Delta\mathcal{B}=2})^{5}%
\sim10^{-18}~(1~$TeV$/\Lambda_{\Delta\mathcal{B}=2})^{5}$ using $\Lambda
_{QCD}\approx300~$MeV~\cite{Rinaldi:2018osy}. In the non-relativistic limit, a two-state Schrodinger-like representation can be constructed, and a standard quantum mechanical calculation expresses the $n-\bar{n}$ oscillation probability as
\begin{equation}
P_{n\rightarrow\bar{n}}(t)=e^{-\Gamma t}\frac{\varepsilon^{2}}{(\Delta
E/2)^{2}+\varepsilon^{2}}\sin^{2}\left(  \sqrt{(\Delta E/2)^{2}+\varepsilon
^{2}}t\right)  \ ,\label{Intro2}%
\end{equation}
where $\Delta E=E_{n}-E_{n^{\mathrm{C}}}$, and $\Gamma$ is the neutron decay width. The CPT theorem predicts $m_{n}=m_{n^{\mathrm{C}}}$, but even a minuscule magnetic field generates $\Delta E\gg\varepsilon$ because of the neutron magnetic moment $\mu(n)=-\mu(n^{\mathrm{C}})=-6.03\times10^{-2}~\mu
$eV/T~\cite{ParticleDataGroup:2024cfk}. Experiments target the so-called quasi-free condition, with $\Delta Et\lesssim1$. Efficient magnetic shielding is then required to allow $t$ to reach about a second, in which case $P_{n\rightarrow\bar{n}}\approx\varepsilon^{2}t^{2}\approx(t/\tau_{osc})^{2}$.
The best limit is quite old~\cite{Baldo-Ceolin:1989vpk,Baldo-Ceolin:1994hzw}, standing at $\tau_{osc}>0.86\times10^{8}$~$s$ using a flight-time of about $0.1~s$, and translates as $\varepsilon<0.8\times10^{-23}~$eV.

The starting point of the present analysis is to observe that if the mixing term oscillates,
\begin{equation}
\varepsilon(t)=\varepsilon_{0}\sin(\omega t)\ ,\label{Intro3}%
\end{equation}
then the oscillation probability becomes
\begin{equation}
P_{n\rightarrow\bar{n}}(t)=e^{-\Gamma t}\frac{\varepsilon_{0}^{2}}%
{((\omega-\Delta E)/2)^{2}+\varepsilon_{0}^{2}}\sin^{2}\left(  \sqrt
{((\omega-\Delta E)/2)^{2}+\varepsilon_{0}^{2}}t\right)  \ .\label{Rabi}%
\end{equation}
The system exhibits a Rabi resonance~\cite{Rabi:1937dgo} for a fixed magnetic
field, when the energy difference matches the mixing term frequency. Compared
to the much weaker enhancements achievable with oscillating magnetic
fields~\cite{Pusch:1982ps,Krstic:1988ix}, it could even in principle reach a
maximal value of $1/2$, turning half the neutrons into antineutrons!

Such an extreme signal is not realistic though. To see this, let us imagine
that $\omega=1\ \mu$eV, and that no external magnetic field is turned on
(except the tiny ambient one). Then, $\omega\gg\Delta E$, and the quasifree
condition would require way too short flight time. Instead, $P_{n\rightarrow
\bar{n}}(t)$ oscillates in the GHz range. For the typical 0.1~s used to derive
the limit for constant $\varepsilon$, it would average to $P_{n\rightarrow
\bar{n}}(t\approx0.1~$s$)\approx\varepsilon_{0}^{2}/\omega^{2}$, which
translates into the constraint $\varepsilon_{0}\lesssim10^{-15}~$eV, eight
orders of magnitude weaker than for a constant $\varepsilon(t)=\varepsilon
_{0}$. If one could turn on an external magnetic field to get precisely at the
Rabi resonance, then $P_{n\rightarrow\bar{n}}(t)\approx e^{-\Gamma t}\sin
^{2}\left(  \varepsilon_{0}t\right)  $. With $\varepsilon_{0}$ as large as
$10^{-15}$~eV, the oscillation characteristic time is about a second, and a
huge number of antineutrons would appear.

The issue though is how close to the resonance one could get. In principle, we
need $\varepsilon_{0}\gg\omega-\Delta E$, with $\omega$ several orders of
magnitude greater than $\varepsilon_{0}$. In practice, adjusting $\Delta E$
requires adjusting the magnetic field, and there is a limit to how precise
this can be done. One way forward would be to repeat the analysis with a
slowly varying or oscillating external magnetic field, but this would bring us
too far afield. Keeping with the simple picture, the important point is that
if we could (optimistically) adjust to a $10^{-6}$ precision the magnetic
field, the oscillation probability would be enhanced by $10^{12} $ since
$P_{n\rightarrow\bar{n}}(t\approx0.1~$s$)\approx\varepsilon_{0}^{2}%
/(\omega-\Delta E)^{2}$. So, even if we do not expect abundant antineutrons to
emerge from a neutron beam, there is ample room for improving the current best
limit on the oscillatory scenario.

The next question is how could the mixing term depend on time. This is where
we connect with our original motivation to entangle $\Delta\mathcal{B}=2$
interactions with dark matter. Indeed, if dark matter is made of scalar fields
carrying $\mathcal{B}=-2$, then at the effective level, we can construct the
coupling
\begin{equation}
\lambda\phi\bar{n}^{\mathrm{C}}n\ .\label{IntroC1}%
\end{equation}
Provided $\phi$ is stable and very light, the dark matter could be made of a
coherent, classical mode $\phi(t)=\phi_{0}\sin(m_{\phi}t)$.
Neutron-antineutron oscillations would then resonate according to
Eq.~(\ref{Rabi}) with $\omega=m_{\phi}$ the scalar mass and $\varepsilon
_{0}=\phi_{0}\lambda$. Setting quite conservatively $\lambda\approx
(\Lambda_{QCD}/\Lambda_{\Delta\mathcal{B}=2})^{6}$, and from $\phi_{0}%
=\sqrt{2\rho_{DM}}/m_{\phi}$ with $\rho_{DM}=0.4\ $GeV$/cm^{3}~$%
\cite{Catena:2009mf}, the parameter $\varepsilon_{0}$ scales like%
\begin{equation}
\varepsilon_{0}=10^{-18}~\text{eV~}\left(  \frac{1~\text{TeV}}{\Lambda
_{\Delta\mathcal{B}=2}}\right)  ^{6}\left(  \frac{1~\mu\text{eV}}{m_{\phi}%
}\right)  \ ,\label{Intro4}%
\end{equation}
which should be within reach experimentally. Notice that the probability
$P_{n\rightarrow\bar{n}}$ scales like $m_{\phi}^{-3}$, so the accessible
$\Lambda_{\Delta\mathcal{B}=2}$ scales increase with decreasing $m_{\phi}$.
This remains true until $\omega-\Delta E$ falls below about $10^{-15}~$eV,
since then, with $\omega-\Delta E>\varepsilon_{0}$, one would recover a quasifree situation, with $P_{n\rightarrow\bar{n}}\approx\varepsilon_{0}%
^{2}t^{2}$. Importantly though, this situation is achieved not thanks to a
magnetic shielding to reduce $\Delta E$, but rather by matching the magnetic
field to the mass of the dark matter particle.

The possibility of having dark matter charged under $\mathcal{B}$ is not new
(see e.g. Ref.~\cite{Heeck:2020nbq} and references there). With $\mathcal{B}%
=2$, typical signature would come from $nn\rightarrow invisibles$, for which
constraints have been obtained and will be improved in the
future~\cite{KamLAND:2005pen,SNO:2022trz,JUNO:2024pur,Dev:2022jbf}. Typically
though, the corresponding scale $\Lambda_{\Delta\mathcal{B}=2}$ must simply be
above the TeV, compatible with Eq.~(\ref{Intro4}). Let us stress that the
dark-matter couplings we consider here are tiny, and become observable only by
exploiting a possible resonant effect. The situation appears similar
concerning possible limits from neutron
stars~\cite{Motta:2018rxp,Berryman:2022zic,Ellis:2018bkr,McKeen:2020oyr}.
Those are known to poorly probe $n-\bar{n}$
oscillations~\cite{Goldman:2024yoh}, but to strongly constrain a possible dark
decay mode for the
neutron~\cite{McKeen:2018xwc,Baym:2018ljz,Cline:2018ami,Bastero-Gil:2024kjo}.
This is a different signature, motivated by the dark matter interpretation of
the persistent neutron lifetime anomaly~\cite{Fornal:2018eol}, on which the
present scenario does not add anything given the size of the coupling in
Eq.~(\ref{Intro4}) and the dark matter mass range considered. Though further
investigations may prove neutron stars to be competitive also for
$\Delta\mathcal{B}=2$ transitions, for the time being, the $n-\bar{n}$
signature discussed here appears more promising.

Now, in a spirit of minimality, it is tempting to identify this scalar field
with the QCD axion. In that case, one would simultaneously solve the strong CP
puzzle~(see e.g. Ref.~\cite{Smith:2024uer} for a short review), identify the
nature of dark matter, and bring in some $\mathcal{B}$ violation, opening new
paths to explain the baryon asymmetry of the Universe. Coincidentally, given
that $\mu(n)=-6.03\times10^{-2}~\mu$eV/T, the preferred axion mass range and
the energy splitting achievable with realistic magnetic fields happen to
overlap, so it is in principle possible to get to the Rabi resonance. The main
goal for the present paper is thus to investigate whether the axion could
induce resonant $n-\bar{n}$ oscillations.

Answering this question is more complicated than it seems. At first sight,
being a pseudoscalar, the only possible couplings of the axion should be
either%
\begin{equation}
a\bar{n}^{\mathrm{C}}\gamma^{5}n\ \ ,\ \ \partial_{\mu}a\bar{n}^{\mathrm{C}%
}\gamma^{\mu}\gamma^{5}n\ .\label{IntroC2}%
\end{equation}
Together with $\phi\bar{n}^{\mathrm{C}}n$, these two close the set of possible
$\Delta\mathcal{B}=2$ fermion bilinears since $\bar{n}^{\mathrm{C}}\gamma
^{\mu}n=\bar{n}^{\mathrm{C}}\sigma^{\mu\nu}n=0$. But a moment of thought shows
that this cannot be the full story:

\begin{itemize}
\item The axion is a Goldstone boson. It must be possible to use the
equivalence theorem to relate the two couplings in Eq.~(\ref{IntroC2}).
Further, it must be possible to understand the derivative one, which is
shift-symmetric, in terms of the PQ symmetry currents. But, realistic axion
$\Delta\mathcal{B}=2$ couplings require first to merge the PQ symmetry with
baryon number, in which case we would rather expect the axion to couple to the
baryon number current, $\partial_{\mu}a\bar{n}\gamma^{\mu}n$. A first question
is thus how is this coupling related to those in Eq.~(\ref{IntroC2}).

\item Again because it is a Goldstone boson, the axion must actually couple
through $(1+ia/v+...)$ combinations, with $v$ the PQ breaking scale. The axion
couplings of Eq.~(\ref{IntroC2}) should thus be accompanied by $\Delta
\mathcal{B}=2$ mass terms (breakings the conservation of the baryon number
current), from which they are relatively suppressed by the large $v$ scale. A
way to evade this pattern must be found otherwise axion-induced neutron
oscillations would always be surpassed by those in vacuum.

\item Technically, the couplings in Eq.~(\ref{IntroC2}) upset the usual
non-relativistic reduction to a Schrodinger-like equation for $n-\bar{n}$
oscillations. In the Dirac representation, $\gamma^{5}$ couples small and
large spinor components, in contrast to $\varepsilon$ in Eq.~(\ref{Intro1})
which does not.\ This is immediately problematic for a $\bar{n}^{\mathrm{C}%
}\gamma^{5}n$ coupling, and raises questions about the use of the equation of
motion to simplify the $\partial_{\mu}a\bar{n}^{\mathrm{C}}\gamma^{\mu}%
\gamma^{5}n$ coupling.
\end{itemize}

Our analysis is organized as follows. In the next Section, we will start by
presenting generic baryonic axion models, elucidating the role of the baryon
number current. In Sec.~3, we tackle the diagonalization of fully general
neutron mass terms, essentially following
Refs.~\cite{Berezhiani:2015uya,Fujikawa:2016sft,Berezhiani:2018xsx}. This is
rather technical, but instrumental to quantify the impact of this
diagonalization first on the electromagnetic and weak interactions of the
neutron in Sec.~4, and then on the derivative interaction terms for the axion
in Sec.~5. There, the anomalous nature of the diagonalization will also be
explored, explaining how the parametrization equivalence between both
couplings of Eq.~(\ref{IntroC2}) works in practice. With all this at hand, we
will go back in Sec.~6 to the initial question of whether it is possible to
induce $n-\bar{n}$ oscillations using the dark matter axion flux. We will find
that though it is possible to turn off oscillations in vacuum, additional
suppressions compared to Eq.~(\ref{Intro4}) nevertheless occur. Observing
axionic $n-\bar{n}$ oscillation at that level would be in conflict with
resolving the strong CP puzzle, thus could only occur for axion-like
particles. This is summarized in our Conclusions, along with perspectives and
possible paths for further improvements.

\section{Baryonic axion models}

Axion models are constructed by first adding a complex scalar field $\phi$,
along with the $U(1)$ symmetry $\phi\rightarrow\exp(i\alpha)\phi$ called the
Peccei-Quinn symmetry (PQ)~\cite{Peccei:1977hh, Peccei:1977ur}. The next
ingredients are first some $U(1)$ invariant couplings to colored fermions,
ensuring the $U(1)$ symmetry becomes anomalous, and second, a spontaneous
symmetry breaking inducing potential $V(\phi^{\dagger}\phi)$. With this, the
axion is the associated Goldstone boson~\cite{Weinberg:1977ma, Wilczek:1977pj},
and it ends up coupled to $G_{\mu\nu}\tilde{G}^{\mu\nu}$, ensuring a
resolution to the strong CP puzzle.

Since baryon number is a symmetry of the SM, it is often possible to add
couplings of $\phi$ to some $\mathcal{B}$-charged combination of
fields~\cite{Quevillon:2020hmx}. At the UV, to maintain renormalizability,
this requires additional colored particles, e.g. diquarks, but we will not be
concerned by the specifics (explicit examples of UV models are presented in
Ref.~\cite{Arias-Aragon:2022byr}). The important point is that this unifies
the $U(1)$ symmetry carried by $\phi$ with baryon number. Both sum up to a
single resulting $U(1)$ that gets spontaneously broken. In our case, we assume
the specific $\mathcal{B}$-charged combination of fields to which $\phi$
couples forces it to carry $\mathcal{B}=-2$.

An effective representation for this construction at the low scale corresponds
to%
\begin{align}
\mathcal{L}_{eff} &  =\partial_{\mu}\phi^{\dagger}\partial^{\mu}\phi
+V(\phi^{\dagger}\phi)+\bar{n}(i\!\not\! \partial-m_{D})n\nonumber\\
&  \ \ \ \ -\frac{1}{2}\varepsilon_{L}\phi\bar{n}_{L}^{\mathrm{C}}n_{L}%
-\frac{1}{2}\varepsilon_{L}^{\ast}\phi^{\dagger}\bar{n}_{L}n_{L}^{\mathrm{C}%
}-\frac{1}{2}\varepsilon_{R}\phi^{\dagger}m_{R}\bar{n}_{R}n_{R}^{\mathrm{C}%
}-\frac{1}{2}\varepsilon_{R}^{\ast}\phi\bar{n}_{R}^{\mathrm{C}}n_{R}%
\ .\label{Bmod1}%
\end{align}
Its only global $U(1)$ is baryon number, and $\phi$ breaks it spontaneously.
Of course, $\varepsilon_{L,R}$ are not to be expected of $\mathcal{O}(1)$
since they encode all the heavy degrees of freedom required to effectively
couple $\phi$ to six-quark states. Rather, if these new states have masses of
$\mathcal{O}(\Lambda)$, then $\varepsilon_{L,R}$ should naively scale as
$\Lambda_{\Delta\mathcal{B}=2}^{-6}$.

The spontaneous symmetry breaking (SSB) is assume to occur at some very high energy, so the true
effective Lagrangian is obtained by expressing $\phi=(\sigma+v)\exp(ia/v)$ and
integrating $\sigma$ out. Setting $m_{L}=\varepsilon_{L}v$ and $m_{R}%
=\varepsilon_{R}v$, we find%
\begin{align}
\mathcal{L}_{eff}  &  =\frac{1}{2}\partial_{\mu}a\partial^{\mu}a+\bar{n}(i
\!\not\! \partial-m_{D})n\nonumber\\
&  \ \ \ \ -\frac{1}{2}m_{L}e^{ia/v}\bar{n}_{L}^{\mathrm{C}}n_{L}-\frac{1}%
{2}m_{L}^{\ast}e^{-ia/v}\bar{n}_{L}n_{L}^{\mathrm{C}}-\frac{1}{2}%
m_{R}e^{-ia/v}m_{R}\bar{n}_{R}n_{R}^{\mathrm{C}}-\frac{1}{2}m_{R}^{\ast
}e^{ia/v}\bar{n}_{R}^{\mathrm{C}}n_{R}\ ,\label{Bmod2}%
\end{align}
or in Dirac form,
\begin{align}
\mathcal{L}_{eff}  &  =\frac{1}{2}\partial_{\mu}a\partial^{\mu}a+\bar{n}(i%
\!\not\!\partial-m_{D})n\nonumber\\
&  \ \ \ \ -\frac{1}{2}m_{S}e^{ia/v}\bar{n}^{\mathrm{C}}n-\frac{1}{2}%
m_{S}^{\ast}e^{-ia/v}\bar{n}n^{\mathrm{C}}-\frac{1}{2}m_{P}e^{-ia/v}\bar
{n}i\gamma^{5}n^{\mathrm{C}}-\frac{1}{2}m_{P}^{\ast}e^{ia/v}\bar
{n}^{\mathrm{C}}i\gamma^{5}n\ ,\label{AxBmod}%
\end{align}
with%
\begin{equation}
m_{S}=\frac{m_{R}^{\ast}+m_{L}}{2},\ \ m_{P}=i\frac{m_{R}-m_{L}^{\ast}}%
{2}\ .\label{msp}%
\end{equation}

Note well that the axion is still assumed to solve the strong CP puzzle.
Alongside the $\Delta\mathcal{B}=2$ sector, both
DFSZ~\cite{Dine:1981rt,Zhitnitsky:1980tq} or KSVZ~\cite{Kim:1979if,
Shifman:1979if} realizations can be added
separately~\cite{Arias-Aragon:2022byr}.\ For example, adopting the latter
approach, we could add to the above model a heavy colored fermion chirally
coupled to $\phi$. This would ensure the necessary $aG^{\mu\nu}\tilde{G}%
_{\mu\nu}$ coupling arises at the low scale.

At this stage, it is conventional to perform a fermionic reparametrization.
Let us first recall how this proceeds in standard axion models. Those starts
from Lagrangians of the type:%
\begin{equation}
\mathcal{L}_{UV}^{chiral}=\partial_{\mu}\phi^{\dagger}\partial^{\mu}%
\phi+V(\phi^{\dagger}\phi)+\bar{\psi}(i%
\!\not\!\partial)\psi+(y \phi\bar{\psi}_{L}\psi_{R}+h.c.)\ .
\end{equation}
The PQ symmetry requires $\psi_{L}$ and $\psi_{R}$ to have different charges,
so its defining characteristic is to be chiral, but it does not involve baryon
or lepton number. After SSB and at low energy, plugging in $\phi\rightarrow
v\exp(ia/v)$, the effective Lagrangian in the so-called exponential
parametrization is, with $m_D = v y$,
\begin{equation}
\mathcal{L}_{eff}^{chiral}=\frac{1}{2}\partial_{\mu}a\partial^{\mu}a+\bar
{\psi}(i\!\not\! \partial)\psi+m_{D}\bar{\psi}\exp(i\gamma^{5}a/v)\psi\ .\label{ChEff1}%
\end{equation}
The fermion fields still transform under the PQ symmetry. The derivative representation of the axion Lagrangian is reached via the fermionic reparametrization $\psi\rightarrow\exp(i\gamma^{5}a/2v)\psi$:%
\begin{equation}
\mathcal{L}_{eff}^{chiral}=\frac{1}{2}\partial_{\mu}a\partial^{\mu}a+\bar
{\psi}(i\!\not\!\partial-m_{D})\psi+\frac{\partial_{\mu}a}{2v}\bar{\psi}\gamma^{\mu}\gamma
^{5}\psi+\mathcal{L}_{Jac}\ .\label{ChEff2}%
\end{equation}
Though not manifest in the Feynman rules, such a change of variable is
allowed, as all observables have been proven to be
identical~\cite{Chisholm:1961tha,Kamefuchi:1961sb}. The terms in
$\mathcal{L}_{Jac}$ break the Goldstone boson shift symmetry%
\begin{equation}
\mathcal{L}_{Jac}=\frac{a}{16\pi^{2}}\left(  g_{Y}B_{\mu\nu}\tilde{B}^{\mu\nu
}+g_{L}W_{\mu\nu}\tilde{W}^{\mu\nu}+g_{G}G_{\mu\nu}\tilde{G}^{\mu\nu}\right)
\ ,
\end{equation}
where $g_{F,W,G}$ depend on the SM and PQ charges of the fermion field $\psi$.
They arise since the PQ symmetry is, by construction, anomalous, and so is the
reparametrization. The presence of these anomalous terms is
important~\cite{Quevillon:2019zrd,Bonnefoy:2020gyh,Quevillon:2021sfz}: they
ensure the whole theory stays anomaly-free. Indeed, in shifting from
pseudoscalar to axial interactions, the axion to gauge-boson triangle graphs
become anomalous. This piece is not physical, and precisely compensated by the
local terms in $\mathcal{L}_{Jac}$.

Returning to the baryonic axion model of Eq.~(\ref{AxBmod}), we can remove
entirely the axion from the Majorana mass terms via the purely baryonic
reparametrization $n\rightarrow\exp(-ia/2v)n$, $n^{\mathrm{C}}=\exp
(ia/2v)n^{\mathrm{C}}$:%
\begin{align}
\mathcal{L}_{eff}  &  =\frac{1}{2}\partial_{\mu}a\partial^{\mu}a+\bar{n}(i%
\!\not\!\partial-m_{D})n+\frac{\partial_{\mu}a}{2v}\bar{n}\gamma^{\mu}n\nonumber\\
&  \ \ \ \ -\frac{1}{2}m_{L}\bar{n}_{L}^{\mathrm{C}}n_{L}-\frac{1}{2}%
m_{L}^{\ast}\bar{n}_{L}n_{L}^{\mathrm{C}}-\frac{1}{2}m_{R}\bar{n}_{R}%
n_{R}^{\mathrm{C}}-\frac{1}{2}m_{R}^{\ast}\bar{n}_{R}^{\mathrm{C}}%
n_{R}+\mathcal{L}_{Jac}\ ,\label{Laxion0}%
\end{align}
with
\begin{equation}
\mathcal{L}_{Jac}=-\frac{a}{16\pi^{2}}\left(  \frac{1}{2}g^{2}W_{\mu\nu}%
^{i}\tilde{W}^{i,\mu\nu}-\frac{1}{2}g^{\prime2}B_{\mu\nu}\tilde{B}^{\mu\nu
}\right)  \ .\label{LjacB}%
\end{equation}
As in the chiral case, a derivative coupling appears to the associated
symmetry current, which is now the usual baryon number current. The anomalous
terms in $\mathcal{L}_{Jac}$ have to be added since the baryon number $U(1)$
is anomalous. If $a\rightarrow W^{+}W^{-}$ is calculated, the triangle graph
from the derivative coupling $\partial_{\mu}a\bar{n}\gamma^{\mu}n$ has an
anomaly that precisely cancel with the local interaction from $\mathcal{L}%
_{Jac}$, ensuring the whole process is anomaly-free and matches the result one
would get using instead the Lagrangian in the exponential parametrization,
Eq.~(\ref{AxBmod}).

\section{Neutron mass eigenstates and standard basis}

In the derivative representation, the axion leaves behind fully generic
neutron mass terms. Indeed, a generic $\Delta\mathcal{B}=2$ free Lagrangian
would take the form%
\begin{align}
\mathcal{L}_{free} &  =\bar{n}_{R}i\!\not\!\partial n_{R}+\bar{n}_{L}i\!\not\!
\partial n_{L}-m_{D}\bar{n}_{R}n_{L}-m_{D}^{\ast}\bar{n}_{L}n_{R}\nonumber\\
&  -\frac{1}{2}m_{L}\bar{n}_{L}^{\mathrm{C}}n_{L}-\frac{1}{2}m_{L}^{\ast}%
\bar{n}_{L}n_{L}^{\mathrm{C}}-\frac{1}{2}m_{R}\bar{n}_{R}n_{R}^{\mathrm{C}%
}-\frac{1}{2}m_{R}^{\ast}\bar{n}_{R}^{\mathrm{C}}n_{R}\ .\label{Lfree}%
\end{align}
In Dirac form, this corresponds to including scalar and pseudoscalar Dirac and
Majorana mass terms,%
\begin{align}
\mathcal{L}_{free} &  =\bar{n}(i\!\not\!\partial)n-\operatorname{Re}m_{D}
\bar{n}n-\operatorname{Im}m_{D}\bar{n}%
i\gamma^{5}n\nonumber\\
&  -\frac{1}{2}m_{S}\bar{n}^{\mathrm{C}}n-\frac{1}{2}m_{S}^{\ast}\bar
{n}n^{\mathrm{C}}-\frac{1}{2}m_{P}\bar{n}i\gamma^{5}n^{\mathrm{C}}-\frac{1}%
{2}m_{P}^{\ast}\bar{n}^{\mathrm{C}}i\gamma^{5}n\ .\label{Lfree2}%
\end{align}
Notice that naively, the CP-conserving limit of Eq.~(\ref{Lfree}) would
prescribe $m_{D,L,R}$ to be real since $(\bar{n}_{R,L}n_{R,L}^{\mathrm{C}%
})^{\dagger}=\bar{n}_{R,L}^{\mathrm{C}}n_{R,L}$, in which case $m_{S}$ and
$m_{D}$ are real, but $m_{P}$ is purely imaginary. The goal of the present
section is to diagonalize these mass terms, or more precisely, to show that no
matter the initial parameters, a change of basis always permits to reach%
\begin{equation}
\mathcal{L}_{free}=\bar{n}(i\!\not\!
\partial)n-m\bar{n}n-\frac{1}{2}\varepsilon(\bar{n}^{\mathrm{C}}n+\bar
{n}n^{\mathrm{C}})\ ,
\end{equation}
for some real parameter $m$ and $\varepsilon$. This is not a new result (see
e.g. Ref.~\cite{Berezhiani:2015uya}), but we will need to set up this
diagonalization procedure in details to study its impact on other sectors, and
in particular on axion couplings.

\subsection{Two-component representation and Takagi's factorization}

To proceed, it is useful to switch to a two-component notation:
\begin{subequations}
\label{WeylSpinor}%
\begin{align}
n  &  =\left(
\begin{array}
[c]{c}%
n_{L}=\xi_{\alpha}\\
n_{R}=\eta^{\dot{\alpha}}%
\end{array}
\right)  \ ,\ \ \bar{n}=\left(
\begin{array}
[c]{cc}%
\bar{n}_{R}=\eta^{\dagger\alpha} & \bar{n}_{L}=\xi_{\dot{\alpha}}^{\dagger}%
\end{array}
\right)  \ ,\ \\
n^{\mathrm{C}}  &  =\left(
\begin{array}
[c]{c}%
n_{R}^{\mathrm{C}}=\eta_{\beta}^{\dagger}\\
n_{L}^{\mathrm{C}}=\xi^{\dagger\dot{\beta}}%
\end{array}
\right)  \ ,\ \bar{n}^{\mathrm{C}}=\left(
\begin{array}
[c]{cc}%
\bar{n}_{L}^{\mathrm{C}}=\xi^{\beta} & \bar{n}_{R}^{\mathrm{C}}=\eta
_{\dot{\beta}}%
\end{array}
\right)  \ .
\end{align}
Some details about our conventions are given in Appendix~\ref{appConv} (also
see e.g. Ref.~\cite{Dreiner:2008tw} for a review). In terms of the $\eta$ and
$\xi$ Weyl spinors, the Lagrangian written in matrix form is
\end{subequations}
\begin{align}
\mathcal{L}  &  =\frac{1}{2}\left[  \left(
\begin{array}
[c]{cc}%
\xi_{\dot{\alpha}}^{\dagger} & \eta_{\dot{\alpha}}%
\end{array}
\right)  \cdot i(\bar{\sigma}^{\mu})^{\dot{\alpha}\alpha}\partial_{\mu}\left(
\begin{array}
[c]{c}%
\xi_{\alpha}\\
\eta_{\alpha}^{\dagger}%
\end{array}
\right)  +\left(
\begin{array}
[c]{cc}%
\xi^{\alpha} & \eta^{\dagger\alpha}%
\end{array}
\right)  \cdot i(\sigma^{\mu})_{\alpha\dot{\alpha}}\partial_{\mu}\left(
\begin{array}
[c]{c}%
\xi^{\dagger\dot{\alpha}}\\
\eta^{\dot{\alpha}}%
\end{array}
\right)  \right] \nonumber\\
&  -\frac{1}{2}\left[  \left(
\begin{array}
[c]{cc}%
\xi^{\alpha} & \eta^{\dagger\alpha}%
\end{array}
\right)  \cdot\left(
\begin{array}
[c]{cc}%
m_{L} & m_{D}\\
m_{D} & m_{R}%
\end{array}
\right)  \cdot\left(
\begin{array}
[c]{c}%
\xi_{\alpha}\\
\eta_{\alpha}^{\dagger}%
\end{array}
\right)  +\left(
\begin{array}
[c]{cc}%
\xi_{\dot{\alpha}}^{\dagger} & \eta_{\dot{\alpha}}%
\end{array}
\right)  \cdot\left(
\begin{array}
[c]{cc}%
m_{L}^{\ast} & m_{D}^{\ast}\\
m_{D}^{\ast} & m_{R}^{\ast}%
\end{array}
\right)  \cdot\left(
\begin{array}
[c]{c}%
\xi^{\dagger\dot{\alpha}}\\
\eta^{\dot{\alpha}}%
\end{array}
\right)  \right]  \ .
\end{align}
To reach this form, the kinetic terms are first split using $\bar{n}(i \!\not\!
\partial)n=\bar{n}^{\mathrm{C}}(i\!\not\!\partial)n^{\mathrm{C}}$.

The four-component spinor made of $\xi$ and $\eta^{\dagger}$ is neither a
Dirac state nor a Majorana state. However, being of the same chirality, those
two Weyl spinors can mix into mass eigenstates. Specifically, the kinetic
terms are invariant under `flavor' $U(2)$ rotations in the space spanned by
$\xi_{\alpha}$ and $\eta_{\alpha}^{\dagger}$:%
\begin{equation}
\left(
\begin{array}
[c]{c}%
\xi_{\alpha}\\
\eta_{\alpha}^{\dagger}%
\end{array}
\right)  \rightarrow\left(
\begin{array}
[c]{c}%
\xi_{\alpha}^{\prime}\\
\eta_{\alpha}^{\prime\dagger}%
\end{array}
\right)  =U^{\dagger}\left(
\begin{array}
[c]{c}%
\xi_{\alpha}\\
\eta_{\alpha}^{\dagger}%
\end{array}
\right)  \ ,\ \ \left(
\begin{array}
[c]{c}%
\xi^{\dagger\dot{\alpha}}\\
\eta^{\dot{\alpha}}%
\end{array}
\right)  \rightarrow\left(
\begin{array}
[c]{c}%
\xi^{\prime\dagger\dot{\alpha}}\\
\eta^{\prime\dot{\alpha}}%
\end{array}
\right)  =U^{T}\left(
\begin{array}
[c]{c}%
\xi^{\dagger\dot{\alpha}}\\
\eta^{\dot{\alpha}}%
\end{array}
\right)  \ .\label{SpinorRot}%
\end{equation}
This can be used to bring the mass matrix to a diagonal form. If we explicitly write all the mass terms as modulus times phase,%
\begin{equation}
M=\left(
\begin{array}
[c]{cc}%
m_{L}e^{i\phi_{L}} & m_{D}e^{i\phi_{D}}\\
m_{D}e^{i\phi_{D}} & m_{R}e^{i\phi_{R}}%
\end{array}
\right)  \rightarrow U^{T}\left(
\begin{array}
[c]{cc}%
m_{L}e^{i\phi_{L}} & m_{D}e^{i\phi_{D}}\\
m_{D}e^{i\phi_{D}} & m_{R}e^{i\phi_{R}}%
\end{array}
\right)  U=\left(
\begin{array}
[c]{cc}%
m_{-} & 0\\
0 & m_{+}%
\end{array}
\right)  \equiv\Sigma\ ,\label{Umass}%
\end{equation}
where $m_{L,R,D}>0$ and $m_{\pm}>0$. From a parameter point of view, $U$ has
four parameters: one angle and three phases.\ Those permit to get rid of the
three phases of $M$, along with a real parameter, leaving $m_{\pm}$ as the two
remaining real parameters~\cite{Berezhiani:2015uya}.

This procedure is not a true diagonalization but rather a Takagi's
factorization, which expresses a symmetric complex matrix $M$ as $\Sigma
=U^{T}MU$ with $\Sigma$ diagonal and positive-definite and $U$ unitary. This
is a consequence of the singular-value decomposition $\Sigma=W^{T}MV$ with $W
$ and $V$ unitary and $\Sigma$ real diagonal and positive semidefinite. With
in addition $M$ symmetric, $V$ and $W$ at most differ by a diagonal matrix of
phases, $V=WP^{\dagger}$, and defining either $U=VP^{-1/2}$ or $U=WP^{\ast
-1/2}$ gives $\Sigma=U^{T}MU$, as announced.

The diagonal entries in $\Sigma$ are the absolute values of the eigenvalues,
equal to the positive square root of the (real) eigenvalues of $M^{\dagger}M$.
When $\phi_{L,R,D}=0$ and with $m_{D}\gg m_{L,R}$, these are
\begin{equation}
m_{\pm}=\frac{1}{2}\left(  \sqrt{4m_{D}^{2}+(m_{L}-m_{R})^{2}}\pm(m_{R}%
+m_{L})\right)  \ ,
\end{equation}
while the $U$ matrix is then a simple $O(2)$ rotation of angle $\alpha$ with
$\tan2\alpha=-|m_{L}-m_{R}|/m_{D}$. For $M$ complex, the eigenvalues can
easily be worked out, but the expression of $U$ is way more complicated
because of the matrix of phases $P$ discussed above.\ Our goal here is to
construct explicitly a matrix $U$ such that $\Sigma=U^{T}MU$ with $\Sigma$
diagonal and positive-definite. A generic $U$ matrix can be arranged into
$U=U(\alpha,\sigma)\cdot P(\phi_{1},\phi_{2})$ with%
\begin{equation}
U(\alpha,\sigma)=\left(
\begin{array}
[c]{cc}%
\cos\alpha & e^{-i\sigma}\sin\alpha\\
-e^{i\sigma}\sin\alpha & \cos\alpha
\end{array}
\right)  \ ,\ \ P(\phi_{1},\phi_{2})=\left(
\begin{array}
[c]{cc}%
e^{i\phi_{1}} & 0\\
0 & e^{i\phi_{2}}%
\end{array}
\right)  \ .
\end{equation}
We will also distinguish baryonic rephasing $P_{\mathcal{B}}(\phi
)=P(\phi/2,-\phi/2)$ and chiral rephasing $P_{\mathcal{C}}(\phi)=P(\phi
/2,\phi/2)=\exp(i\phi/2)\times\mathbf{1}$. Notice that $U(\alpha
,\sigma)=P_{\mathcal{B}}(-\sigma)\cdot U(\alpha,0)\cdot P_{\mathcal{B}}%
(\sigma)$, $\det U(\alpha,\sigma)=\det P_{\mathcal{B}}(\phi)=1$ but $\det
P_{\mathcal{C}}(\phi)=e^{i\phi}$. In practice, we will construct $U$ in steps
as a product of unitary matrices assuming the above decomposition, following
essentially the procedure outlined in Ref.~\cite{Fujikawa:2016sft}.

\subsection{Step-wise diagonalization in the general case}

As a first step, it is convenient to get rid of some phases using
$P_{\mathcal{B}}$ and $P_{\mathcal{C}}$ transformations. First, since
\begin{equation}
\arg\det M=k\pi+\arctan\left(  \frac{m_{D}^{2}\sin2\phi_{D}-m_{L}m_{R}%
\sin(\phi_{R}+\phi_{L})}{m_{D}^{2}\cos2\phi_{D}-m_{L}m_{R}\cos(\phi_{R}%
+\phi_{L})}\right)  \equiv-2\zeta\ ,\label{ksi}%
\end{equation}
for some integer $k$, we perform a chiral rotation to put the mass matrix into
the form
\begin{equation}
M\overset{P_{\mathcal{C}}(\zeta)}{\rightarrow}\left(
\begin{array}
[c]{cc}%
m_{L}e^{i(\phi_{L}-\phi_{D})} & m_{D}\\
m_{D} & m_{R}e^{i(\phi_{R}-\phi_{D})}%
\end{array}
\right)  e^{i\zeta+\phi_{D}}\ ,
\end{equation}
which it has a real positive determinant (when $m_{D}>m_{R,L}$). Since
$\det\Sigma$ is also real positive, $U$ must now be an $SU(2)$ matrix (up to
its sign which is ambiguous), i.e., a combination of $U(\alpha,\sigma)$ and
$P_{\mathcal{B}}(\phi)$ only. We can also remove a further phase $\phi
_{\Delta}=(\phi_{R}-\phi_{L})/2$, leaving only the common phase $\phi_{\Sigma
}=(\phi_{R}+\phi_{L}-2\phi_{D})/2$ as
\begin{equation}
M\overset{P_{\mathcal{B}}(\phi_{\Delta})}{\rightarrow}\left(
\begin{array}
[c]{cc}%
m_{L}e^{i\phi_{\Sigma}} & m_{D}\\
m_{D} & m_{R}e^{i\phi_{\Sigma}}%
\end{array}
\right)  e^{i\zeta+\phi_{D}}\ .
\end{equation}

In view of the mass terms, a good strategy is to perform yet another
$P_{\mathcal{B}}$ rephasing to force $m_{R}^{\ast}-m_{L}$ or $m_{R}^{\ast
}+m_{L} $ to be phase-free. Choosing the former, the target form is%
\begin{equation}
M\overset{P_{\mathcal{B}}(\beta)}{\rightarrow}\left(
\begin{array}
[c]{cc}%
m_{L}e^{i(\phi_{\Sigma}+\beta)} & m_{D}\\
m_{D} & m_{R}e^{i(\phi_{\Sigma}-\beta)}%
\end{array}
\right)  e^{i\zeta+\phi_{D}}\equiv\left(
\begin{array}
[c]{cc}%
\varepsilon_{s}e^{i\delta}-\varepsilon_{p} & m_{D}\\
m_{D} & \varepsilon_{s}e^{-i\delta}+\varepsilon_{p}%
\end{array}
\right)  e^{i\zeta+\phi_{D}}\ ,\label{Form1}%
\end{equation}
with
\begin{equation}
\varepsilon_{p}=\frac{1}{2}\sqrt{m_{L}^{2}-2m_{L}m_{R}\cos(2\phi_{\Sigma
})+m_{R}^{2}}=|m_{P}|\ ,\ \ \varepsilon_{s}=\frac{1}{2}\sqrt{m_{L}^{2}%
+2m_{L}m_{R}\cos(2\phi_{\Sigma})+m_{R}^{2}}=|m_{S}|\ .
\end{equation}
To ensure that $\varepsilon_{p}$ is positive, $\beta$ must be taken in
specific quadrants depending on that of $\phi_{\Sigma}$, thereby fixing that
of $\delta$. If by definition $\arctan(x)\in\lbrack-\pi/2,\pi/2]$, the compact
representation
\begin{equation}
\beta=\frac{\pi}{2}s_{\phi_{\Sigma}}+\arctan\left(  \frac{1}{\tan\phi_{\Sigma
}}\frac{m_{L}-m_{R}}{m_{L}+m_{R}}\right)  \ ,\ \delta=\frac{\pi}{2}%
s_{2\phi_{\Sigma}}+\arctan\left(  \frac{m_{R}^{2}-m_{L}^{2}}{2m_{L}m_{R}%
\sin2\phi_{\Sigma}}\right)  \ ,
\end{equation}
ensures the correct signs with $s_{\phi}=\operatorname*{sign}(\sin\phi)$,
whatever $\phi_{\Sigma}$ and the sign of $m_{L}-m_{R}$.

The interest of the form of Eq.~(\ref{Form1}) is that an $O(2)$ rotation
$U(\alpha,0)$ now permits to move $\varepsilon_{p}$ to the Dirac mass terms,
with $\tan2\alpha=-\varepsilon_{p}/m_{D}$, with $\alpha\in\lbrack-\pi/4,0]$
when $\varepsilon_{p}$ is smaller than $m_{D}$. This rotation is the core of
the diagonalization procedure; this is where states get mixed with their
complex conjugate. In some sense, at the level of the neutron Dirac spinor, it
can be viewed as a kind of Bogoliubov transformation~\cite{Fujikawa:2016sft}.
It brings the mass matrix in the form%
\begin{equation}
M\overset{U(\alpha,0)}{\rightarrow}\left(
\begin{array}
[c]{cc}%
\varepsilon_{s}\sqrt{1-r_{\alpha}^{2}}e^{i\eta} & \sqrt{m_{D}^{2}%
+\varepsilon_{p}^{2}}+i\varepsilon_{s}r_{\alpha}\\
\sqrt{m_{D}^{2}+\varepsilon_{p}^{2}}+i\varepsilon_{s}r_{\alpha} &
\varepsilon_{s}\sqrt{1-r_{\alpha}^{2}}e^{-i\eta}%
\end{array}
\right)  e^{i\zeta+\phi_{D}},\ \eta=\frac{\pi}{2}s_{2\phi_{\Sigma}}%
-\arctan\left(  \frac{1}{\tan\delta\cos2\alpha}\right)  ,
\end{equation}
with $r_{\alpha}=\sin\delta\sin2\alpha$.

From here, only trivial transformations remain. The phase $\eta$ can be
removed with $P_{\mathcal{B}}(-\eta)$, leaving both identical diagonal and
off-diagonal entries, and a $45%
%TCIMACRO{\U{b0}}%
%BeginExpansion
{{}^\circ}%
%EndExpansion
$ rotation suffices to reach%
\begin{equation}
M\overset{P_{\mathcal{B}}(-\eta)\cdot U(\pi/4,0)}{\rightarrow}\left(
\begin{array}
[c]{cc}%
\varepsilon_{s}\sqrt{1-r_{\alpha}^{2}}-\sqrt{m_{D}^{2}+\varepsilon_{p}^{2}%
}-i\varepsilon_{s}r_{\alpha} & 0\\
0 & \varepsilon_{s}\sqrt{1-r_{\alpha}^{2}}+\sqrt{m_{D}^{2}+\varepsilon_{p}%
^{2}}+i\varepsilon_{s}r_{\alpha}%
\end{array}
\right)  e^{i\zeta+\phi_{D}}\ .
\end{equation}
Since $\zeta$ ensures that the mass matrix has unit determinant, the phases of
both entries are equal and opposite, and can be eliminated with
$P_{\mathcal{B}}(\gamma)$:%
\begin{equation}
M\overset{P_{\mathcal{B}}(\gamma)}{\rightarrow}\left(
\begin{array}
[c]{cc}%
-m_{-} & 0\\
0 & m_{+}%
\end{array}
\right)  \ ,\ \ \gamma=-\frac{1}{2}\arctan\left(  \frac{2r_{\alpha}%
\sqrt{1-r_{\alpha}^{2}}\varepsilon_{s}^{2}}{m_{D}^{2}+\varepsilon_{p}%
^{2}-(1-2r_{\alpha}^{2})\varepsilon_{s}^{2}}\right)  \ ,
\end{equation}
where%
\begin{equation}
m_{\pm}=\sqrt{m_{D}^{2}+\varepsilon_{p}^{2}+\varepsilon_{s}^{2}\pm
2\varepsilon_{s}\sqrt{m_{D}^{2}+\varepsilon_{p}^{2}}\cos\delta}\ \ .
\end{equation}
A final rephasing with $P(\pi/2,0)$ makes the first entry positive.
Alternatively, the \textit{standard basis} relevant for neutron oscillation is
reached by undoing the previous $45 {{}^\circ}$ rotation%
\begin{equation}
M\overset{U(-\pi/4,0)}{\rightarrow}\left(
\begin{array}
[c]{cc}%
\varepsilon & m\\
m & \varepsilon
\end{array}
\right)  \ ,\ m=\frac{m_{+}+m_{-}}{2}\ ,\ \ \varepsilon=\frac{m_{+}-m_{-}}%
{2}>0\ .\label{basis}%
\end{equation}

\subsection{Approximate expressions and special cases}

Altogether, the transformation matrix can be expressed in various forms, for
example as
\begin{subequations}
\label{TotU}%
\begin{align}
U &  =P_{\mathcal{B}}(\beta+\phi_{\Delta})\cdot U(\alpha,0)\cdot
P_{\mathcal{B}}(-\eta)\cdot U(\pi/4,0)\cdot P_{\mathcal{B}}(\gamma)\cdot
U(-\pi/4,0)\cdot P_{\mathcal{C}}(\zeta)\\
&  =P_{\mathcal{B}}(\phi_{\Delta})\cdot U(\alpha,-\beta)\cdot P_{\mathcal{B}%
}(\beta-\eta)\cdot U(\gamma/2,\pi/2)\cdot P_{\mathcal{C}}(\zeta)\ .
\end{align}
Notice that as constructed here, $U$ depends on many parameters, yet only four
should be sufficient since $U\in U(2)$. However, because of the complex
dependencies on the initial mass matrix parameters, and of the non-linear
composition rule for unitary matrices, there does not seem to be a simple way
to express this redundancy analytically. To proceed, let us instead take the
route of the mass expansion.

Numerically, given that $m_{D}\gg m_{L,R}$, some approximate expressions for
$U$ can be constructed. All but $\beta$ and $\delta$ admit $1/m_{D}$
expansions:
\end{subequations}
\begin{subequations}
\label{mDexp}%
\begin{align}
\alpha &  =-\frac{\varepsilon_{p}}{2m_{D}}+\frac{\varepsilon_{p}^{3}}%
{6m_{D}^{3}}+\mathcal{O}(m_{D}^{-5})\ ,\\
\eta &  =\delta+\frac{m_{R}^{2}-m_{L}^{2}}{8\varepsilon_{s}^{2}}\frac
{m_{L}m_{R}\sin2\phi_{\Sigma}}{2m_{D}^{2}}+\mathcal{O}(m_{D}^{-4})\ ,\\
\gamma &  =-\frac{2\varepsilon_{s}}{m_{D}}\frac{m_{L}m_{R}\sin2\phi_{\Sigma}%
}{2m_{D}^{2}}+\mathcal{O}(m_{D}^{-5})\ ,\\
\zeta &  =-\phi_{D}+\frac{m_{L}m_{R}\sin2\phi_{\Sigma}}{2m_{D}^{2}%
}+\mathcal{O}(m_{D}^{-4})\ .
\end{align}
For completeness, the standard parameters in the mass matrix, Eq.~(\ref{basis}), have the expansion
\end{subequations}
\begin{equation}
m=m_{D}+\frac{\varepsilon_{p}^{2}}{2m_{D}}+\mathcal{O}(m_{D}^{-3}%
)\ ,\ \ \ \varepsilon=\varepsilon_{s}-\frac{m_{L}^{2}m_{R}^{2}\sin^{2}%
(2\phi_{\Sigma})}{8\varepsilon_{s}m_{D}^{2}}+\mathcal{O}(m_{D}^{-3}%
)\ .\label{mDexp2}%
\end{equation}
Three observations can be made on the basis of these expansions.

First, $\gamma$ starts at $\mathcal{O}(m_{D}^{-3})$. For all practical
purpose, it can be set to zero. This simplifies the $U$ matrix to%
\begin{equation}
U=P_{\mathcal{B}}(\phi_{\Delta})\cdot U(\alpha,-\beta)\cdot P_{\mathcal{B}%
}(\beta-\eta)\cdot P_{\mathcal{C}}(\zeta)+\mathcal{O}(m_{D}^{-3})\ .
\end{equation}
Notice how setting $\gamma$ to zero removes both $\pi/4$ rotations in $U$.
This is a first instance where one can see the special properties of the basis
of Eq.~(\ref{basis}), that will be apparent again when discussing current transformations.

Second, in view of the occurrence of $\sin2\phi_{\Sigma}$ in all these
expansions, both $\phi_{\Sigma}=0+k\pi$ and $\phi_{\Sigma}=\pi/2+k\pi$ appear
as interesting limiting cases. Setting $\phi_{\Sigma}=0$ or $\pi/2$ makes
$m_{L,R}$ purely real or imaginary, respectively (but for a global phase).
Correspondingly, $m_{S}$ is real and $m_{P}$ imaginary for $\phi_{\Sigma}=0$,
and the opposite holds at $\phi_{\Sigma}=\pi/2$. These two situations
correspond to interchanging $\varepsilon_{s}$ and $\varepsilon_{p} $ since
(all angles given up to $k\pi$)%
\[%
\begin{tabular}
[c]{cc}\hline
$\phi_{\Sigma}\rightarrow0$ & $\phi_{\Sigma}\rightarrow\pi/2$\\\hline
$\beta\rightarrow0$ & $\beta\rightarrow\pi/2$\\
\multicolumn{2}{c}{$\delta,\eta,\gamma\rightarrow0,\zeta\rightarrow-\phi_{D}$%
}\\
\multicolumn{2}{c}{$\varepsilon\rightarrow\varepsilon_{s}$}\\
$\varepsilon_{s}\rightarrow(m_{L}+m_{R})/2$ & $\varepsilon_{s}\rightarrow
|m_{L}-m_{R}|/2$\\
$\varepsilon_{p}\rightarrow|m_{L}-m_{R}|/2$ & $\varepsilon_{p}\rightarrow
(m_{L}+m_{R})/2$\\\hline
\end{tabular}
\
\]

Third, the two scenarios above are actually quite different. Technically, this
is apparent if we now also send $m_{L}-m_{R}$ to zero. Care is needed to do
that because not all the angles stay well-defined. This is reminiscent of the
usual breakdown of the polar representation when the modulus of a complex
number goes to zero.\ In practice, it shows up as poles in $1/\varepsilon_{p}$
for $\phi_{\Sigma}\rightarrow0$, or $1/\varepsilon_{s}$ for $\phi_{\Sigma
}\rightarrow\pi/2$, in
\begin{subequations}
\begin{align}
\sin\delta &  =\frac{m_{L}m_{R}\sin2\phi_{\Sigma}}{2\varepsilon_{p}%
\varepsilon_{s}}\ ,\ \ \cos\delta=\frac{m_{R}^{2}-m_{L}^{2}}{4\varepsilon
_{p}\varepsilon_{s}}\ ,\\
\sin\beta &  =\frac{m_{L}+m_{R}}{2\varepsilon_{p}}\sin\phi_{\Sigma}%
\ ,\ \ \cos\beta=\frac{m_{R}-m_{L}}{2\varepsilon_{p}}\cos\phi_{\Sigma}\ .
\end{align}

The combined limit $\phi_{\Sigma}\rightarrow0$ together with $m_{L}%
-m_{R}\rightarrow0$ can be made stable by writing the $U$ matrix as
\end{subequations}
\begin{equation}
U=P_{\mathcal{B}}(\phi_{\Delta})\cdot U(\alpha,-\beta)\cdot P_{\mathcal{B}%
}(\beta-\delta)\cdot P_{\mathcal{B}}(\delta-\eta)\cdot P_{\mathcal{C}}%
(\zeta)+\mathcal{O}(m_{D}^{-3})\ .\label{Limit0}%
\end{equation}
Indeed, the combination $\eta-\delta$ is obviously regular, see
Eq.~(\ref{mDexp}), the combination $\beta-\delta$ satisfies
\begin{equation}
\sin(\beta-\delta)=\frac{m_{R}-m_{L}}{2\varepsilon_{s}}\sin\phi_{\Sigma
}\ ,\ \ \cos(\beta-\delta)=\frac{m_{R}+m_{L}}{2\varepsilon_{s}}\cos
\phi_{\Sigma}\ ,\label{bmd}%
\end{equation}
while the $1/\varepsilon_{p}$ pole cancels out in the product $e^{\pm i\beta
}\sin\alpha$ occurring in the off-diagonal entries of $U(\alpha,-\beta)$. All
in all, $P_{\mathcal{B}}(\beta-\delta)$, $P_{\mathcal{B}}(\delta-\eta)$, and
$U(\alpha,-\beta)$ all become trivial when $\phi_{\Sigma}$ and $m_{L}-m_{R}$
go to zero (up to irrelevant global signs). This is expected since at that
point, the mass matrix is immediately in the standard form of Eq.~(\ref{basis}%
), so no rephasings nor Bogoliubov rotations are needed. In
App.~\ref{ExplicitU}, we show the explicit form of $U$, using the
decomposition of Eq.~(\ref{Limit0}).

The situation is not the same for $\phi_{\Sigma}\rightarrow\pi/2$ because
$\varepsilon_{s}$ defines the angle $\delta$ (and thus $\eta$) via the
expression $\varepsilon_{s}e^{i\delta}$, see Eq.~(\ref{Form1}). Adjusting the
relative speed at which $\phi_{\Sigma}\rightarrow\pi/2$ and $m_{L}\rightarrow
m_{R}$, $\delta$ (or $\eta$) can take any value one wishes. One may think this
is an artifact of the parametrization and that it should be possible to
somehow interchange $\varepsilon_{p}$ and $\varepsilon_{s}$, but this is not
so. These two parameters do not play symmetrical roles, as is most evident the
expansion of the mixing term $\varepsilon$, Eq.~(\ref{mDexp2}). This shows
that actually, $\phi_{\Sigma}=\pi/2$ and $m_{L}=m_{R}$ is very
special~\cite{Fujikawa:2016sft}: it is the only point in parameter space where
$\varepsilon\rightarrow0$, that is, where an effective baryon number is
restored (besides the trivial $m_{L}=m_{R}=0$). This is the point where the
mass matrix takes the form, with $\bar{m}\equiv m_{L}=m_{R}$:%
\begin{equation}
M=\left(
\begin{array}
[c]{cc}%
i\bar{m} & m_{D}\\
m_{D} & i\bar{m}%
\end{array}
\right)  \ ,\label{SpuriousB0}%
\end{equation}
which is diagonalized by $U=U(\alpha,\pi/2)$ with $\tan2\alpha=\bar{m}/m_{D}$.
The parameter $\varepsilon=0$ because the two eigenvalues are identical, as
trivially follows from $M^{\dagger}M=(m_{D}^{2}+\bar{m}^{2})\mathbf{1}$.
Physically, this means that the Lagrangian%
\begin{equation}
\mathcal{L}_{free}=\bar{n}(i\!\not\!
\partial)n-m_{D}\bar{n}n+\frac{1}{2}\bar{m}(\bar{n}i\gamma^{5}n^{\mathrm{C}%
}+\bar{n}^{\mathrm{C}}i\gamma^{5}n)\ ,\label{SpuriousB}%
\end{equation}
actually do not induce any $\Delta\mathcal{B}$ effects, being equivalent to%
\begin{equation}
\mathcal{L}_{free}=\bar{n}(i\!\not\!%
\partial)n-\sqrt{m_{D}^{2}+\bar{m}^{2}}\bar{n}n\ .
\end{equation}
We will see in the following that though $\Delta\mathcal{B}\rightarrow0$ in
the free Lagrangian, some imprints of $U(\alpha,\pi/2)$ can remain in several
other sectors which develop $\Delta\mathcal{B}=2$ components.

\section{Electroweak neutron interactions in the standard basis}

In the SM, the neutron has two main electroweak interactions: the magnetic
moment and weak $\beta$ decay. The associated observables are crucial to
define the CP-conserving spinor basis, and to differentiate neutrons from
antineutrons. At the same time, these interactions are not preserved under the
diagonalization of the mass term, so it is important to understand how the two
play together.

\subsection{Magnetic and electric dipole moments}

Let us first consider the electromagnetic couplings%
\begin{equation}
\mathcal{L}_{em}=-\frac{\mu}{4}\bar{n}_{R}\sigma^{\mu\nu}n_{L}F_{\mu\nu}%
-\frac{\mu^{\ast}}{4}\bar{n}_{L}\sigma^{\mu\nu}n_{R}F_{\mu\nu}=-\frac
{\operatorname{Re}\mu}{2}\bar{n}\sigma^{\mu\nu}nF_{\mu\nu}+i\frac
{\operatorname{Im}\mu}{2}\bar{n}\sigma^{\mu\nu}\gamma^{5}nF_{\mu\nu}\ ,
\end{equation}
where we identify the neutron anomalous magnetic moment $a_{n}$ (MDM) and
electric dipole moment $d_{n}$ (EDM) as%
\begin{equation}
\operatorname{Re}\mu=\frac{ea_{n}}{2m_{D}}\ ,\ \ d_{n}=\operatorname{Im}\mu\ .
\end{equation}
The quark model gives $a_{n}=-2$, in fairly good agreement with the measured
$a_{n}\approx-1.913$~\cite{ParticleDataGroup:2024cfk}. On the other hand,
$d_{n}$ is experimentally tiny, $d_{n}<1.8\times10^{-26}$ e$cm~$%
\cite{Abel:2020pzs}.

Written in terms of Weyl spinors, $\mathcal{L}_{em}$ takes the form%
\begin{align}
\mathcal{L}_{em} &  =\frac{F_{\mu\nu}}{2}\left[
\begin{array}
[c]{c}%
\left(
\begin{array}
[c]{cc}%
\xi^{\alpha} & \eta^{\dagger\alpha}%
\end{array}
\right) \\
~
\end{array}
\left(
\begin{array}
[c]{cc}%
0 & -1\\
1 & 0
\end{array}
\right)  \mu(\sigma^{\mu\nu})_{\alpha}^{\,\,\,\beta}\left(
\begin{array}
[c]{c}%
\xi_{\beta}\\
\eta_{\beta}^{\dagger}%
\end{array}
\right)  \ \ \right. \nonumber \\
&  \ \ \ \ \ \ \ \ \ \left.  -%
\begin{array}
[c]{c}%
\left(
\begin{array}
[c]{cc}%
\xi_{\dot{\alpha}}^{\dagger} & \eta_{\dot{\alpha}}%
\end{array}
\right)  \\ 
~
\end{array}
\left(
\begin{array}
[c]{cc}%
0 & -1\\
1 & 0
\end{array}
\right)  \mu^{\ast}(\bar{\sigma}^{\mu\nu})_{\,\,\,\dot{\beta}}^{\dot{\alpha}%
}\left(
\begin{array}
[c]{c}%
\xi^{\dagger\dot{\beta}}\\
\eta^{\dot{\beta}}%
\end{array}
\right)  \right]  \ .
\end{align}
The antisymmetry in flavor space means that the magnetic moment transforms in
a very simple way under $U(2)$:
\begin{equation}
\mu\rightarrow\mu^{\prime}=\mu\det(U)=\mu e^{i\zeta}\ ,
\end{equation}
with $\zeta$ given in Eq.~(\ref{ksi}) and its expansion in Eq.~(\ref{mDexp}).
Thus, only the chiral part of the $U(2)$ transformation contributes, while
neither the baryonic rephasing nor the Bogoliubov rotation affect the
electromagnetic couplings. If only the Dirac mass term has a phase, then
$\zeta=-\phi_{D}$ and we recover the well-known fact that a eliminating a
pseudoscalar mass term via a chiral rotation redefines the splitting between
MDM and EDM (see e.g. the discussion in Ref.~\cite{Smith:2023htu} and
references there). In view of this, conventionally, the phase of $\mu$ is
chosen such that no EDM arises in the basis in which the Dirac mass term is
real. Thus, we set%
\[
\mu=|\mu|e^{i\phi_{D}}\ .
\]
As a result, under the action of $U$, the shift is $\mu\rightarrow\mu^{\prime
}=|\mu|\exp(i(\zeta+\phi_{D}))$. The change in $a_{n}$ is then driven by that
of $m_{D}\rightarrow m$ with $m$ in Eq.~(\ref{basis}), with the result
\begin{subequations}
\label{EDMMDM}%
\begin{align}
a_{n}^{\prime} &  =\frac{2m}{e}|\mu|\cos(\zeta+\phi_{D})=a_{n}\left(
1+\frac{\varepsilon_{p}^{2}}{2m_{D}^{2}}\right)  +\mathcal{O}(m_{D}%
^{-3})\ ,\ \ \\
d_{n}^{\prime} &  =|\mu|\sin(\zeta+\phi_{D})=|\mu|\frac{m_{L}m_{R}\sin
2\phi_{\Sigma}}{2m_{D}^{2}}+\mathcal{O}(m_{D}^{-3})\approx\frac{ea_{n}}%
{2m_{D}}\frac{m_{L}m_{R}\sin2\phi_{\Sigma}}{2m_{D}^{2}}\ .
\end{align}
If we assume $m_{L}\approx m_{R}$ and $\phi_{\Sigma}\sim\mathcal{O}(1)$, then
the current bound on $d_{n}$ implies $m_{L,R}/m_{D}\lesssim10^{-6}$, far less
constraining than the bound derived from neutron-antineutron oscillations,
$\varepsilon\sim m_{L,R}<0.80\times10^{-23}~$eV.

\subsection{Weak interaction and beta decay}
\label{weak}

At the level of the proton and neutron, the weak interaction vertex is
\end{subequations}
\begin{align}
\mathcal{L}_{weak}  &  =\frac{g}{2\sqrt{2}}V_{ud}\{\bar{p}\gamma^{\mu}%
(g_{V}+g_{A}\gamma_{5})n\}W_{\mu}^{+}+h.c.\nonumber\\
&  =\frac{g}{4\sqrt{2}}V_{ud}\{\bar{p}\gamma^{\mu}(g_{V}+g_{A}\gamma
_{5})n-\bar{n}^{\mathrm{C}}\gamma^{\mu}(g_{V}-g_{A}\gamma_{5})p^{\mathrm{C}%
}\}W_{\mu}^{+}+h.c.\ ,\label{Weak}%
\end{align}
where $g$ is the weak gauge coupling, $V_{ud}$ is the CKM matrix element,
$g_{V}$ and $g_{A}$ are the vector and axial vector form-factors for which we
neglect the momentum dependence. Because of the conserved vector current
hypothesis (or QED current conservation), $g_{V}$ stays very close to one even
at the hadronic level, but $g_{A}$ receives large corrections. Experimentally,
$\lambda=g_{A}/g_{V}=-1.2756(13)$~\cite{ParticleDataGroup:2024cfk}. This
assumes $g_{A,V}$ are real, as is the case in the SM. If they are not, their
relative phase can be extracted from a specific T-odd triple correlation in
the decay of polarized neutrons. Current experimental searches sum up to $\arg
g_{A}/g_{V}=180.017(26) {{}^\circ}$, compatible with $\pi$~\cite{ParticleDataGroup:2024cfk}.

The $U$ transformation acts differently on $n_{R}$ and $n_{L}$, and thus
affects $\lambda$ and its phase. Given Eq.~(\ref{WeylSpinor}) and
Eq.~(\ref{SpinorRot}), we have%
\begin{subequations}
\begin{align}
n  &  =\left(
\begin{array}
[c]{c}%
n_{L}=\xi_{\alpha}\\
n_{R}=\eta^{\dot{\alpha}}%
\end{array}
\right)  \rightarrow n^{\prime}=\left(
\begin{array}
[c]{c}%
U_{11}^{\ast}\xi_{\alpha}+U_{21}^{\ast}\eta_{\alpha}^{\dagger}\\
U_{12}\xi^{\dagger\dot{\alpha}}+U_{22}\eta^{\dot{\alpha}}%
\end{array}
\right)  =\left(
\begin{array}
[c]{c}%
U_{11}^{\ast}n_{L}+U_{21}^{\ast}n_{R}^{\mathrm{C}}\\
U_{12}n_{L}^{\mathrm{C}}+U_{22}n_{R}%
\end{array}
\right)  \ ,\\
n^{\mathrm{C}}  &  =\left(
\begin{array}
[c]{c}%
n_{R}^{\mathrm{C}}=\eta_{\beta}^{\dagger}\\
n_{L}^{\mathrm{C}}=\xi^{\dagger\dot{\beta}}%
\end{array}
\right)  \rightarrow n^{\prime\mathrm{C}}=\left(
\begin{array}
[c]{c}%
U_{12}^{\ast}\xi_{\beta}+U_{22}^{\ast}\eta_{\beta}^{\dagger}\\
U_{11}\xi^{\dagger\dot{\beta}}+U_{21}\eta^{\dot{\beta}}%
\end{array}
\right)  =\left(
\begin{array}
[c]{c}%
U_{12}^{\ast}n_{L}+U_{22}^{\ast}n_{R}^{\mathrm{C}}\\
U_{11}n_{L}^{\mathrm{C}}+U_{21}n_{R}%
\end{array}
\right)  \ ,
\end{align}
\end{subequations}
which can be written as
\begin{subequations}
\label{Fullnnc}%
\begin{align}
n  &  \rightarrow n^{\prime}=\left(
\begin{array}
[c]{cc}%
U_{11}^{\ast} & 0\\
0 & U_{22}%
\end{array}
\right)  n+\left(
\begin{array}
[c]{cc}%
U_{21}^{\ast} & 0\\
0 & U_{12}%
\end{array}
\right)  n^{\mathrm{C}}=\left(
\begin{array}
[c]{cc}%
\det U^{\ast} & 0\\
0 & 1
\end{array}
\right)  \left(  U_{22}n+U_{12}\gamma^{5}n^{\mathrm{C}}\right)  \ ,\\
n^{\mathrm{C}}  &  \rightarrow n^{\prime\mathrm{C}}=\left(
\begin{array}
[c]{cc}%
U_{22}^{\ast} & 0\\
0 & U_{11}%
\end{array}
\right)  n^{\mathrm{C}}+\left(
\begin{array}
[c]{cc}%
U_{12}^{\ast} & 0\\
0 & U_{21}%
\end{array}
\right)  n=\left(
\begin{array}
[c]{cc}%
1 & 0\\
0 & \det U
\end{array}
\right)  \left(  U_{22}^{\ast}n^{\mathrm{C}}-U_{12}^{\ast}\gamma^{5}n\right)
\ ,
\end{align}
since for a $2\times2$ unitary matrix, $U_{11}=U_{22}^{\ast}\det U$ and
$U_{12}=-U_{21}^{\ast}\det U$. The weak vertices in the standard basis are
then obtained by substituting these expressions in Eq.~(\ref{Weak}). Thus, only
$\det U$ can affect the phase of $\lambda$ since a pure baryonic rephasing,
$P_{\mathcal{B}}(\phi)$, only enters in $U_{22}$ and can always be compensated
by a corresponding baryonic rephasing of the proton (both have $\mathcal{B}%
=1$). By contrast, the chiral rephasing tuned by $\det U$ cannot be
compensated if the proton already has its usual real Dirac mass term. The
$n^{\mathrm{C}}$ component of $n$ is tuned by $U_{12}$, which is always
suppressed by $m_{L,R}/m_{D}$. The presence of $\gamma_{5}$ is important
though. It comes from the Bogoliubov-like rotation $U=U(\alpha,0)$, for which
\end{subequations}
\begin{equation}
n\rightarrow n^{\prime}=n\cos\alpha+\gamma^{5}n^{\mathrm{C}}\sin
\alpha\ ,\ n^{\mathrm{C}}\rightarrow n^{\prime\mathrm{C}}=n^{\mathrm{C}}%
\cos\alpha-\gamma^{5}n\sin\alpha\ .\label{BogoN}%
\end{equation}
This shows that $g_{V}$ and $g_{A}$ get mixed under $U(\alpha,0)$. This will
be particularly relevant when discussing anomalies in Sec.~\ref{anom}.

Let us be more explicit and calculate the variation of $\lambda$. We can write%
\begin{equation}
\bar{p}\gamma^{\mu}(g_{V}+g_{A}\gamma_{5})n\rightarrow|U_{22}|\bar{p}%
\gamma^{\mu}(\det U^{\ast}(g_{V}-g_{A})P_{L}+(g_{V}+g_{A})P_{R})n+...\ ,
\end{equation}
which amounts to%
\begin{equation}
\left(
\begin{array}
[c]{c}%
g_{V}\\
g_{A}%
\end{array}
\right)  \rightarrow\frac{1}{2}|U_{22}^{\ast}|\left(
\begin{array}
[c]{cc}%
1+\det U^{\ast} & 1-\det U^{\ast}\\
1-\det U^{\ast} & 1+\det U^{\ast}%
\end{array}
\right)  \left(
\begin{array}
[c]{c}%
g_{V}\\
g_{A}%
\end{array}
\right)  \ .
\end{equation}
Thus, the ratio $\lambda=g_{A}/g_{V}$ for neutron mass eigenstates (or
equivalently for neutron fields in the standard basis) is different than for
weak interaction eigenstates. It is given by%
\begin{equation}
\arg\lambda=\arg\frac{g_{A}}{g_{V}}\rightarrow\arctan\left(  \frac{\lambda
^{2}-1}{2\lambda}\operatorname{Im}\det U^{\ast}\right)  \ .
\end{equation}
Phenomenologically, this effect is totally irrelevant given the expected size
of $m_{L,R}/m_{D}$. The interesting point here is rather conceptual:
CP-violation is again tuned entirely by the chiral phase extracted via $\det
U$. Majorana phases do not contribute\footnote{It would be interesting at this
stage to calculate CP-violation in $n-\bar{n} $ oscillations, following
Ref.~\cite{McKeen:2015cuz}. We leave that for future studies.}, as could have
been expected since this observable is again $\Delta\mathcal{B}=0$.

In practice, even if $\operatorname{Im}\det U$ was much larger, it would be
difficult to observe this effect. Neutrons would have to be produced somehow,
say via strong interaction processes. If we model those by some pion
exchanges, starting from $\{\bar{p}\gamma^{5}n\}\pi$ vertices, the same
substitution as in Eq.~(\ref{Fullnnc}) must be done. Moving to the standard
neutron basis also creates $\Delta\mathcal{B}=2$ strong interaction effects.
So, a pure neutron beam would be difficult to isolate.

In this respect, let us stress that even in a scenario in which $\varepsilon
=0$, some $\Delta\mathcal{B}=2$ effects can remain because the strong and weak
vertices depend on the whole $U$ matrix. In other words, if one is looking for
antineutrons in a neutral beam obtained from some nuclear reactions, there will
be some antineutron pollution in the beam due to the $\Delta\mathcal{B}=2$
strong interactions, and some of the neutrons will decay to antiprotons due to
the $\Delta\mathcal{B}=2$ weak vertices.\ Of course, this is not truly an
oscillation, and it does not depend on time or energy, but it must in
principle be taken into account.

Let us illustrate this in the $\varepsilon=0$ scenario, assuming phases cancel
out leaving just $U=U(\alpha,0)$ with $\tan2\alpha=-\bar{m}/m_{D}$, $\bar
{m}=m_{L}=m_{R}$. In some sense, the wrong $\mathcal{B}$ state pollution at
production and decay is maximal in this case. At production, the neutral beam
is made of the $n^{\prime}$ states of Eq.~(\ref{BogoN}). This state is a mass
eigenstate that propagates and then decays, with its true neutron and
antineutron components producing $n\rightarrow\cos\alpha\left\vert pe^{-}%
\nu\right\rangle +\sin\alpha\left\vert \bar{p}e^{+}\nu\right\rangle $ and
$n^{\mathrm{C}}=\left\vert \bar{p}e^{+}\nu\right\rangle \cos\alpha-\sin
\alpha\left\vert pe^{-}\nu\right\rangle $. Thus, we have a probability
$\sin^{2}(2\alpha)$ of detecting a positron even without any $n-\bar{n}$
oscillations. Notice that this effect disappears in the opposite limit in
which $\varepsilon$ is maximal but $\varepsilon_{p}=0$, and thus $\alpha=0$.
This arises for $m_{L}=m_{R}$ with $\phi_{\Sigma}=0$, that is, when the
neutrons are from the start in the standard basis and $U$ is trivial, so
neither the strong nor the weak vertex have $\Delta\mathcal{B}=2 $ components.

\section{Axion derivative interactions and anomalies}

Whether through reparametrization or directly from some UV models, there may
be a series of derivative couplings to neutron pairs. In full generality, four
such currents can be constructed:%
\begin{equation}
\mathcal{L}_{der}=\frac{\partial_{\mu}a}{2v}\left(  \eta_{V}\bar{n}\gamma
^{\mu}n+\eta_{A}\bar{n}\gamma^{\mu}\gamma^{5}n+\lambda_{A}\bar{n}^{\mathrm{C}%
}\gamma^{\mu}\gamma^{5}n+\lambda_{A}^{\ast}\bar{n}\gamma^{\mu}\gamma
^{5}n^{\mathrm{C}}\right)  \ ,
\end{equation}
which can be split into%
\begin{align}
\mathcal{L}_{der} &  =\frac{\partial_{\mu}a}{4v}\eta_{V}(\bar{n}_{L}%
\gamma^{\mu}n_{L}-\bar{n}_{L}^{\mathrm{C}}\gamma^{\mu}n_{L}^{\mathrm{C}}%
+\bar{n}_{R}\gamma^{\mu}n_{R}-\bar{n}_{R}^{\mathrm{C}}\gamma^{\mu}%
n_{R}^{\mathrm{C}})\nonumber\\
&  \ \ \ +\frac{\partial_{\mu}a}{4v}\eta_{A}(\bar{n}_{R}\gamma^{\mu}n_{R}%
-\bar{n}_{R}^{\mathrm{C}}\gamma^{\mu}n_{R}^{\mathrm{C}}-\bar{n}_{L}\gamma
^{\mu}n_{L}+\bar{n}_{L}^{\mathrm{C}}\gamma^{\mu}n_{L}^{\mathrm{C}})\nonumber\\
&  \ \ \ +\frac{\partial_{\mu}a}{2v}(\lambda_{A}(\bar{n}_{L}^{\mathrm{C}%
}\gamma^{\mu}n_{R}-\bar{n}_{R}^{\mathrm{C}}\gamma^{\mu}n_{L})+\lambda
_{A}^{\ast}(\bar{n}_{R}\gamma^{\mu}n_{L}^{\mathrm{C}}-\bar{n}_{L}\gamma^{\mu
}n_{R}^{\mathrm{C}}))\ .
\end{align}
There is no $\Delta\mathcal{B}=2$ vector current because of Eq.~(\ref{DEF2}),
which translates as $\bar{n}^{\mathrm{C}}\gamma^{\mu}n=-\bar{n}^{\mathrm{C}%
}\gamma^{\mu}n$ and $\bar{n}\gamma^{\mu}n^{\mathrm{C}}=-\bar{n}\gamma^{\mu
}n^{\mathrm{C}}$. Switching to a two-component notation, identifying chiral
states with the corresponding Weyl spinor, Eq.~(\ref{WeylSpinor}), these
couplings can be written as%
\begin{equation}
\mathcal{L}_{der}=\frac{\partial_{\mu}a}{4v}\eta^{m}J_{m}^{\mu}\ ,\ \ \ J_{m}%
^{\mu}=\left(
\begin{array}
[c]{cc}%
\xi^{\alpha} & \eta^{\dagger\alpha}%
\end{array}
\right)  \cdot\left(  \sigma^{m}\right)  \cdot(\sigma^{\mu})_{\alpha\dot
{\beta}}\left(
\begin{array}
[c]{c}%
\xi^{\dagger\dot{\beta}}\\
\eta^{\dot{\beta}}%
\end{array}
\right)  \ ,\label{derInt}%
\end{equation}
where $\eta_{0}=\eta_{A}$ tunes the axial current, $\eta_{1}%
=2\operatorname{Re}\lambda_{A}$ and $\eta_{2}=-2\operatorname{Im}\lambda_{A}$
tune the CP-conserving and CP-violating parts of the $\Delta\mathcal{B}=2$
axial current, and $\eta_{3}=-\eta_{V}$ tunes the baryon number vector
current. Summation over the Latin index $m=0,1,2,3$ is Euclidian, so $\eta
_{m}=\eta^{m}$.

The basis of currents is complete in the sense that the full hermitian basis
made of Pauli matrices is present. Given the spinor involved and
Eq.~(\ref{SpinorRot}), the action of $U$ is not the same as on the mass
matrix, but rather
\begin{equation}
\eta_{m}\sigma^{m}\rightarrow\eta_{m}^{\prime}\sigma^{m}=U^{T}\eta_{m}%
\sigma^{m}U^{\ast}\ ,\label{UJ}%
\end{equation}
so that $\eta^{m}J_{m}^{\mu}=\eta^{\prime m}J_{m}^{\prime\mu}$, with
$J_{m}^{\prime\mu}$ expressed as in Eq.~(\ref{derInt}) but in terms of
$\xi^{\prime},\eta^{\prime}$. Unitarity of $U$ implies that $\eta_{0}^{\prime
}=\eta_{0}$ and $\vec{\eta}^{\prime2}=\vec{\eta}^{2}$, $\vec{\eta}=(\eta
_{1},\eta_{2},\eta_{3})$. Thus, the vector and $\Delta\mathcal{B}=2$ axial
currents get mixed, but the $\Delta\mathcal{B}=0$ axial current is totally
unaffected by the diagonalization of the mass matrix. This makes sense since
that current is intrinsically $\Delta\mathcal{B}=0$, while the vector current
is actually the baryon number current. Said differently, the chiral rotation
required to make the Dirac mass term real relies on the axial symmetry, which
is preserved by these currents (up to anomalies, see later).

Here, it is important to define $U$ with respect to the standard basis in
Eq.~(\ref{basis}), and not the basis in which the mass matrix is diagonal.
Indeed, it is only in the former basis that the transformation of the currents
make sense from the point of view of the CP symmetry. The reason is
essentially that $U\rightarrow1$ when $\alpha$ and all the phases go to zero
only in the basis of Eq.~(\ref{basis}). By contrast, denoting $U_{diag}$ that
diagonalizing the mass matrix, we have $U_{diag}\rightarrow U(\pi/4,0)$ in
that same limit. For neutrons, it makes sense to keep the states as close as
possible to Dirac states. Indeed, if one thinks in terms of the
non-relativistic expansion, it is the real Dirac mass term that defines the
basis from which one can identify e.g. the CP-violating EDM term.

The action of $U$ on the four-component coefficients $\eta_{m}$,
Eq.~(\ref{UJ}), reduces to that of an orthogonal rotation on its vector
components,
\begin{equation}
\vec{\eta}\rightarrow\vec{\eta}^{\prime}=\mathcal{R}(U)\cdot\vec{\eta}\ ,
\end{equation}
with
\begin{equation}
\mathcal{R}(U)=\left(
\begin{array}
[c]{ccc}%
1 & 0 & 0\\
0 & c_{\gamma} & -s_{\gamma}\\
0 & s_{\gamma} & c_{\gamma}%
\end{array}
\right)  \cdot\left(
\begin{array}
[c]{ccc}%
c_{\eta} & -s_{\eta} & 0\\
s_{\eta} & c_{\eta} & 0\\
0 & 0 & 1
\end{array}
\right)  \cdot\left(
\begin{array}
[c]{ccc}%
c_{2\alpha} & 0 & s_{2\alpha}\\
0 & 1 & 0\\
-s_{2\alpha} & 0 & c_{2\alpha}%
\end{array}
\right)  \cdot\left(
\begin{array}
[c]{ccc}%
c_{\beta+\phi_{\Delta}} & s_{\beta+\phi_{\Delta}} & 0\\
-s_{\beta+\phi_{\Delta}} & c_{\beta+\phi_{\Delta}} & 0\\
0 & 0 & 1
\end{array}
\right)  \ ,\label{RU}%
\end{equation}
where $c_{\delta},s_{\delta}\equiv\cos\delta,\sin\delta$. It is quite
remarkable that the full-fledged unitary transformation collapses to this
orthogonal form, in terms of the same angles as defined in the diagonalization
procedure. Further, these angles here have simple interpretations, with
$\beta+\phi_{\Delta}$ and $\eta$ inducing CP-violating mixings between the two
$\Delta\mathcal{B}=2$ axial current, $\alpha$ mixing the CP-conserving vector
and $\Delta\mathcal{B}=2$ axial currents, and $\gamma$ mixing the
CP-conserving vector and CP-violating $\Delta\mathcal{B}=2 $ axial current.
Actually, this last rotation is equivalent to%
\begin{equation}
\left(
\begin{array}
[c]{ccc}%
1 & 0 & 0\\
0 & c_{\gamma} & -s_{\gamma}\\
0 & s_{\gamma} & c_{\gamma}%
\end{array}
\right)  =\left(
\begin{array}
[c]{ccc}%
c_{\pi/2} & 0 & -s_{\pi/2}\\
0 & 1 & 0\\
s_{\pi/2} & 0 & c_{\pi/2}%
\end{array}
\right)  \cdot\left(
\begin{array}
[c]{ccc}%
c_{\gamma} & s_{\gamma} & 0\\
-s_{\gamma} & c_{\gamma} & 0\\
0 & 0 & 1
\end{array}
\right)  \cdot\left(
\begin{array}
[c]{ccc}%
c_{\pi/2} & 0 & s_{\pi/2}\\
0 & 1 & 0\\
-s_{\pi/2} & 0 & c_{\pi/2}%
\end{array}
\right)  \ ,
\end{equation}
which corresponds to the $\pi/4$ and $-\pi/4$ rotations in $U$, see
Eq.~(\ref{TotU}). In this form, the baryonic rephasing $P_{\mathcal{B}}(\phi)$
always act on the $\Delta\mathcal{B}=2$ subspace, while $U(\phi,0)$ rotations
always mix the CP-conserving currents only. Notice that in the CP-conserving
limit, all that remain is the angle $\alpha$ that mixes $\eta_{1}$ and
$\eta_{3}$, but leaves the CP-violating $\eta_{2}$ untouched. Any mixing into
$\eta_{2}$ requires CP-violation, and this violation is parametrically
independent of the chiral phase relevant for the EDM term or beta decay, since
$\arg\det U$ cancels out here, see Eq.~(\ref{UJ}).

The explicit expression of $\mathcal{R}(U)$ in terms of the original
parameters of the mass matrix is:%
\begin{equation}
\mathcal{R}(U)=\left(
\begin{array}
[c]{ccc}%
\dfrac{m_{L}\cos\phi_{L}+m_{R}\cos\phi_{R}}{2\varepsilon_{s}} & \dfrac
{m_{R}\sin\phi_{R}-m_{L}\sin\phi_{L}}{2\varepsilon_{s}} & \dfrac{m_{L}%
^{2}-m_{R}^{2}}{4m_{D}\varepsilon_{s}}\\
\dfrac{m_{L}\sin\phi_{L}-m_{R}\sin\phi_{R}}{2\varepsilon_{s}} & \dfrac
{m_{L}\cos\phi_{L}+m_{R}\cos\phi_{R}}{2\varepsilon_{s}} & -\dfrac{m_{R}%
m_{L}\sin2\phi_{\Sigma}}{2m_{D}\varepsilon_{s}}\\
\dfrac{m_{R}\cos\phi_{R}-m_{L}\cos\phi_{L}}{2m_{D}} & \dfrac{m_{R}\sin\phi
_{R}+m_{L}\sin\phi_{L}}{2m_{D}} & 1
\end{array}
\right)  +\mathcal{O}(m_{D}^{-2})\ ,\label{RotR}%
\end{equation}
for $\phi_{D}=0$ (for $\phi_{D}\neq0$, simply replace $\phi_{L,R}%
\rightarrow\phi_{L,R}-\phi_{D}$). One can check that $\mathcal{R}(U)^{T}%
\cdot\mathcal{R}(U)=\mathcal{R}(U)\cdot\mathcal{R}(U)^{T}=1$, up to terms of
$\mathcal{O}(m_{D}^{-2})$. This form shows that the mismatch in the $m_{L}$
and $m_{R}$ phases is the source of an $\mathcal{O}(1)$ reorganization between
CP-conserving and CP-violating component of the $\Delta\mathcal{B}=2$ axial
currents. This is the equivalent in the $\Delta\mathcal{B}$ sector of how the
MDM and EDM get reorganized when the Dirac mass term is made real. As such, it
is worth noting that they are much larger than the CP-violating $\det U$ term,
which is $\mathcal{O}(m_{D}^{-2}) $, see Eq.~(\ref{mDexp}). By contrast, any
mixing between the vector current and the $\Delta\mathcal{B}=2$ axial current
requires a Bogoliubov rotation, and is thus necessarily of $\mathcal{O}%
(m_{D}^{-1})$.

\subsection{Equations of motion and classical divergences}

The most general derivative interactions of a scalar field with Weyl fermion
in Eq.~(\ref{derInt}) can be written in a symmetric way as
\begin{equation}
\mathcal{L}_{der}=\frac{\partial_{\mu}a}{4v}\eta_{m}\left[  \left(
\begin{array}
[c]{cc}%
\xi^{\alpha} & \eta^{\dagger\alpha}%
\end{array}
\right)  \cdot\sigma^{m}\cdot(\sigma^{\mu})_{\alpha\dot{\beta}}\left(
\begin{array}
[c]{c}%
\xi^{\dagger\dot{\beta}}\\
\eta^{\dot{\beta}}%
\end{array}
\right)  -\left(
\begin{array}
[c]{cc}%
\xi_{\dot{\beta}}^{\dagger} & \eta_{\dot{\beta}}%
\end{array}
\right)  \cdot\sigma^{mT}\cdot(\bar{\sigma}^{\mu})^{\dot{\beta}\alpha}\left(
\begin{array}
[c]{c}%
\xi_{\alpha}\\
\eta_{\alpha}^{\dagger}%
\end{array}
\right)  \right]  \ ,
\end{equation}
using $\eta_{\dot{\alpha}}(\bar{\sigma}^{\mu})^{\dot{\alpha}\alpha}\xi
_{\alpha}=-\xi^{\alpha}(\sigma^{\mu})_{\alpha\dot{\alpha}}\eta^{\dot{\alpha}}
$ and similar. Remember that $\eta_{m}$ gives the specific couplings of the
$a$ field to the axial, vector, CP-violating and CP-conserving $\Delta
\mathcal{B}=2$ axial currents. In the presence of several scalar fields,
summation is understood. Integrating by part, the derivative is made to act on
the fermion fields. At that stage, assuming them on-shell, the Weyl equation
of motion simplifies the coupling to combinations of scalar and pseudoscalar couplings.

Specifically, using identities like $\xi_{\dot{\alpha}}^{\dagger}\left(
\bar{\sigma}^{\mu}\right)  ^{\dot{\alpha}\alpha}\partial_{\mu}\xi_{\alpha}%
=\xi^{\alpha}\left(  \sigma^{\mu}\right)  _{\alpha\dot{\alpha}}\partial_{\mu
}\xi^{\dagger\dot{\alpha}}$ up to a total derivative, the four free equations
of motion (EoM) are
\begin{equation}
i(\sigma^{\mu})_{\alpha\dot{\alpha}}\partial_{\mu}\left(
\begin{array}
[c]{c}%
\xi^{\dagger\dot{\alpha}}\\
\eta^{\dot{\alpha}}%
\end{array}
\right)  =\left(
\begin{array}
[c]{cc}%
m_{L} & m_{D}\\
m_{D} & m_{R}%
\end{array}
\right)  \left(
\begin{array}
[c]{c}%
\xi_{\alpha}\\
\eta_{\alpha}^{\dagger}%
\end{array}
\right)  ,\ i(\bar{\sigma}^{\mu})^{\dot{\alpha}\alpha}\partial_{\mu}\left(
\begin{array}
[c]{c}%
\xi_{\alpha}\\
\eta_{\alpha}^{\dagger}%
\end{array}
\right)  =\left(
\begin{array}
[c]{cc}%
m_{L}^{\ast} & m_{D}^{\ast}\\
m_{D}^{\ast} & m_{R}^{\ast}%
\end{array}
\right)  \left(
\begin{array}
[c]{c}%
\xi^{\dagger\dot{\alpha}}\\
\eta^{\dot{\alpha}}%
\end{array}
\right)  \ ,
\end{equation}
the derivative interactions collapses to pseudoscalar and scalar
interactions:
\begin{align}
\mathcal{L}_{der} &  =i\frac{a}{4v}\eta_{m}\left(
\begin{array}
[c]{cc}%
\xi^{\alpha} & \eta^{\dagger\alpha}%
\end{array}
\right)  \cdot\left\{  \sigma^{m}\cdot\left(
\begin{array}
[c]{cc}%
m_{L} & m_{D}\\
m_{D} & m_{R}%
\end{array}
\right)  +\left(
\begin{array}
[c]{cc}%
m_{L} & m_{D}\\
m_{D} & m_{R}%
\end{array}
\right)  \cdot\sigma^{mT}\right\}  \cdot\left(
\begin{array}
[c]{c}%
\xi_{\alpha}\\
\eta_{\alpha}^{\dagger}%
\end{array}
\right) \nonumber\\
&  -i\frac{a}{4v}\eta_{m}\left(
\begin{array}
[c]{cc}%
\xi_{\dot{\beta}}^{\dagger} & \eta_{\dot{\beta}}%
\end{array}
\right)  \cdot\left\{  \left(
\begin{array}
[c]{cc}%
m_{L}^{\ast} & m_{D}^{\ast}\\
m_{D}^{\ast} & m_{R}^{\ast}%
\end{array}
\right)  \cdot\sigma^{m}+\sigma^{mT}\cdot\left(
\begin{array}
[c]{cc}%
m_{L}^{\ast} & m_{D}^{\ast}\\
m_{D}^{\ast} & m_{R}^{\ast}%
\end{array}
\right)  \right\}  \cdot\left(
\begin{array}
[c]{c}%
\xi^{\dagger\dot{\beta}}\\
\eta^{\dot{\beta}}%
\end{array}
\right)  \ .
\end{align}
Notice that this form is compatible with the diagonalization procedure, in the
sense that using Eqs.~(\ref{SpinorRot}) and~(\ref{UJ}),%
\begin{equation}
\mathcal{L}_{der}=i\frac{a}{4v}\eta_{m}^{\prime}\left(
\begin{array}
[c]{cc}%
\xi^{\prime\alpha} & \eta^{\prime\dagger\alpha}%
\end{array}
\right)  \cdot\left\{  \sigma^{m}\cdot\left(
\begin{array}
[c]{cc}%
\varepsilon & m\\
m & \varepsilon
\end{array}
\right)  +\left(
\begin{array}
[c]{cc}%
\varepsilon & m\\
m & \varepsilon
\end{array}
\right)  \cdot\sigma^{mT}\right\}  \cdot\left(
\begin{array}
[c]{c}%
\xi_{\alpha}^{\prime}\\
\eta_{\alpha}^{\prime\dagger}%
\end{array}
\right)  +h.c.\ ,
\end{equation}
where $\eta_{0}^{\prime}=\eta_{0},\ \vec{\eta}^{\prime}=\mathcal{R}%
(U)\cdot\vec{\eta}$ and the primed spinors are those in the standard
basis~(\ref{basis}).

\subsubsection{Axial current}

For the axial current, setting $\eta_{m}\sigma^{m}=\eta_{0}\sigma^{0}$ the
induced couplings are obtained from $\mathcal{L}_{free}$, Eq.~(\ref{Lfree}),
by shifting%
\begin{equation}
\left(
\begin{array}
[c]{cc}%
m_{L} & m_{D}\\
m_{D} & m_{R}%
\end{array}
\right)  \rightarrow\left(  1+i\frac{a}{2v}\right)  \eta_{0}\left(
\begin{array}
[c]{cc}%
m_{L} & m_{D}\\
m_{D} & m_{R}%
\end{array}
\right)  \ .
\end{equation}
With $\eta_{0}=1$, this is consistent with the chiral couplings%
\begin{equation}
\mathcal{L}_{chiral}=\bar{n}(i\!\not\! \partial)n
-m_{D}e^{ia/v}\bar{n}_{R}n_{L}-\frac{1}{2}m_{L}e^{ia/v}\bar{n}%
_{L}^{\mathrm{C}}n_{L}-\frac{1}{2}m_{R}e^{ia/v}\bar{n}_{R}n_{R}^{\mathrm{C}%
}+h.c.\ \ ,
\end{equation}
expanded to linear order in the scalar field $a$. This is expected on the
basis of the usual Ward identity $\partial_{\mu}A^{\mu}=2imP$ with $A^{\mu
}=\bar{n}\gamma^{\mu}\gamma^{5}n$ and $P=\bar{n}\gamma^{5}n$, generalized to
account for the presence of Majorana mass terms as (for $m_{D} $ real):%
\begin{equation}
\partial_{\mu}(\bar{n}\gamma^{\mu}\gamma^{5}n)\overset{\mathrm{classical}}%
{=}2im_{D}\bar{n}\gamma^{5}n+m_{S}^{\ast}\bar{n}i\gamma^{5}n^{\mathrm{C}%
}+m_{S}\bar{n}^{\mathrm{C}}i\gamma^{5}n-m_{P}^{\ast}\bar{n}^{\mathrm{C}%
}n-m_{P}\bar{n}n^{\mathrm{C}}\ ,
\end{equation}
with $m_{S,P}$ defined in Eq.~(\ref{msp}). This last equation can also be
checked directly using the EoM for the four Dirac fields $n$, $n^{\mathrm{C}}
$, $\bar{n}$, and $\bar{n}^{\mathrm{C}}$. Notice also that the scalar field
decouples if it becomes constant, $a(x)\rightarrow a_{0}$. This is manifest in
the derivative representation, and can be reproduced in the $\mathcal{L}_{p}$
representation via the constant chiral rotation $n_{L}\rightarrow
n_{L}e^{ia_{0}/2v}$, $n_{R}\rightarrow n_{R}e^{-ia_{0}/2v}$.

\subsubsection{Vector current}

Similarly, for the vector current $\eta_{m}\sigma^{m}=\eta_{3}\sigma^{3}$, the
induced couplings are obtained from $\mathcal{L}_{free}$, Eq.~(\ref{Lfree}),
by shifting%
\begin{equation}
\left(
\begin{array}
[c]{cc}%
m_{L} & m_{D}\\
m_{D} & m_{R}%
\end{array}
\right)  \rightarrow\left(
\begin{array}
[c]{cc}%
m_{L} & m_{D}\\
m_{D} & m_{R}%
\end{array}
\right)  +\left(  1+i\frac{a}{2v}\right)  \eta_{3}\left(
\begin{array}
[c]{cc}%
m_{L} & 0\\
0 & m_{R}%
\end{array}
\right)  \ .
\end{equation}
To leading order in $a$ with $\eta_{3}=1$, this corresponds to the baryonic
couplings%
\begin{equation}
\mathcal{L}_{baryonic}=\bar{n}(i%
\!\not\!\partial)n-m_{D}\bar{n}_{R}n_{L}-\frac{1}{2}m_{L}e^{ia/v}\bar{n}%
_{L}^{\mathrm{C}}n_{L}-\frac{1}{2}m_{R}e^{-ia/v}\bar{n}_{R}n_{R}^{\mathrm{C}%
}+h.c.\ .
\end{equation}
Again, the decoupling in the limit $a(x)\rightarrow a_{0}$ is consistent since
$a_{0}$ can be removed by the baryonic rephasing $n_{L,R}\rightarrow
n_{L,R}e^{ia_{0}/2v}$. Also, the use of the EoM corresponds to the classical
Ward identity $\partial_{\mu}V^{\mu}=0$ with $V^{\mu}=\bar{n}\gamma^{\mu}n$
generalized to%
\begin{equation}
\partial_{\mu}(\bar{n}\gamma^{\mu}n)\overset{\mathrm{classical}}{=}im_{S}%
\bar{n}^{\mathrm{C}}n-m_{S}^{\ast}i\bar{n}n^{\mathrm{C}}+m_{P}\bar{n}%
\gamma^{5}n^{\mathrm{C}}-m_{P}^{\ast}\bar{n}^{\mathrm{C}}\gamma^{5}n\ .
\end{equation}

\subsubsection{$\Delta\mathcal{B}=2$ currents\label{dbcurr}}

The situation is significantly different for the two $\Delta\mathcal{B}=2$
currents. Setting $\eta_{\nu}\sigma^{\nu}=\eta_{1}\sigma^{1}+\eta_{2}%
\sigma^{2}$, their divergences are%
\begin{equation}
\left(
\begin{array}
[c]{cc}%
m_{L} & m_{D}\\
m_{D} & m_{R}%
\end{array}
\right)  \rightarrow\left(
\begin{array}
[c]{cc}%
m_{L} & m_{D}\\
m_{D} & m_{R}%
\end{array}
\right)  +i\frac{a}{4v}(\eta_{1}+i\eta_{2})\left(
\begin{array}
[c]{cc}%
2m_{D} & m_{L}\\
m_{L} & 2m_{D}%
\end{array}
\right)  +i\frac{a}{4v}(\eta_{1}-i\eta_{2})\left(
\begin{array}
[c]{cc}%
0 & m_{R}\\
m_{R} & 0
\end{array}
\right)  \ .
\end{equation}
The scalar field couplings are no longer aligned with the mass terms. With
Dirac and Majorana mass terms getting mixed up, this cannot correspond to the
linearized version of a simple exponential representation. This is expected
since from the point of view of the possible $U(1)$ symmetries, either one
rotates $n_{L}$ and $n_{R}$ by the same phase, or by opposite phases.\ This
then fixes the rotations of all the other fields via the definition
$n_{L,R}^{\mathrm{C}}=\mathrm{C}\bar{n}_{L,R}^{T}$. These two rephasings
correspond to the chiral and baryonic rephasing, leaving no more room for an
extra $U(1)$. The same is visible in the Weyl representation, where $\xi$ and
$\eta$ can either be rotated by the same phase or opposite phases.

By contrast, the transformations associated to $\eta_{1}$ and $\eta_{2}$
directly affects the very definitions of $\mathrm{C}$ since they correspond to
mixing states with their complex conjugate. Infinitesimally, they act e.g. as
$n_{L,R}^{\mathrm{C}}=\mathrm{C}\bar{n}_{L,R}^{T}+\chi_{L,R}n_{R,L}%
+\mathcal{O}(\chi_{L,R}^{2})$, $\bar{n}_{L,R}^{\mathrm{C}}=-n_{L,R}%
^{T}\mathrm{C}^{\dagger}-\xi_{R,L}\bar{n}_{R,L}+\mathcal{O}(\xi_{R,L}^{2})$,
with $\chi_{L,R}$, $\xi_{R,L}\sim a/v$ up to phases. This corresponds to the
classical equations (for $m_{D}$ real)
\begin{subequations}
\label{dB2eom}%
\begin{align}
&  \partial_{\mu}(\bar{n}\gamma^{\mu}\gamma^{5}n^{\mathrm{C}}+\bar
{n}^{\mathrm{C}}\gamma^{\mu}\gamma^{5}n)\overset{\mathrm{classical}}{=}%
2im_{D}(\bar{n}^{\mathrm{C}}\gamma^{5}n+\bar{n}\gamma^{5}n^{\mathrm{C}%
})+4i\operatorname{Re}m_{S}\bar{n}\gamma^{5}n-4\operatorname{Re}m_{P}\bar
{n}n\ ,\\
&  i\partial_{\mu}(\bar{n}\gamma^{\mu}\gamma^{5}n^{\mathrm{C}}-\bar
{n}^{\mathrm{C}}\gamma^{\mu}\gamma^{5}n)\overset{\mathrm{classical}}{=}%
2m_{D}(\bar{n}^{\mathrm{C}}\gamma^{5}n-\bar{n}\gamma^{5}n^{\mathrm{C}%
})-4i\operatorname{Im}m_{S}\bar{n}\gamma^{5}n-4\operatorname{Im}m_{P}\bar
{n}n\ .
\end{align}

The decoupling in the $a(x)\rightarrow a_{0}$ limit is no longer manifest in
this case, but must nevertheless hold. Consider the situation in which
$m_{L,R}=0$, so baryon number is only broken by the derivative couplings and
must disappear in the decoupling limit. For simplicity, consider also that
only the $J_{1}$ current is present. After integrating by part and using the
EoM, it generates pseudoscalar $\Delta\mathcal{B}=2$ couplings. In the
$a(x)\rightarrow a_{0}$ limit, those become Majorana mass terms, whose form
precisely match that in Eq.~(\ref{SpuriousB}). This means these Majorana terms
are spurious, and baryon number is indeed conserved in the decoupling limit,
as it should. Turning on $J_{2}$ does not change this picture since a rotation
permits to recover $\eta_{1}\sigma^{1}+\eta_{2}\sigma^{2}\rightarrow\eta
_{1}^{\prime}\sigma^{1}$. Note that this brings also a phase in the Dirac mass
term which has to be included in the EoM to get the decoupling. Somewhat
similarly, the decoupling also works in the presence of electroweak couplings.
Technically, this is ensured by the cancellation of two different
contributions. First, one should include the weak vertices in the EoM, and
this generates additional weak couplings in Eqs.~(\ref{dB2eom}). In the
$a(x)\rightarrow a_{0}$ limit, these become $\Delta\mathcal{B}=2$ weak
vertices. Second, the other terms in Eqs.~(\ref{dB2eom}) become mass terms in
the $a(x)\rightarrow a_{0}$ limit that require a non-trivial $U$
transformation, generating again some $\Delta\mathcal{B}=2$ weak vertices that
precisely cancel with those arising from the EoM.

\subsection{Anomalies and local terms}
\label{anom}

As soon as fermion fields undergo $U(1)$ rotations, the question of anomalies
must be posed. In the present context, the $U$ transformation contains both
the chiral and baryonic $U(1)$s, which are anomalous. To proceed, let us first
discuss these anomalies, and their dependence on the fermion basis, and then
calculate how the Lagrangian changes when the mass term is brought into the
standard form.

\subsubsection{Triangle graphs}

The classical divergences presented in the previous section do not survive
quantization, but must be corrected for anomalous terms. Our goal here is to
calculate those from triangle graphs involving one current insertion, and a
pair of charged weak gauge bosons.

\begin{figure}[t]
\centering\includegraphics[width=0.90\textwidth]{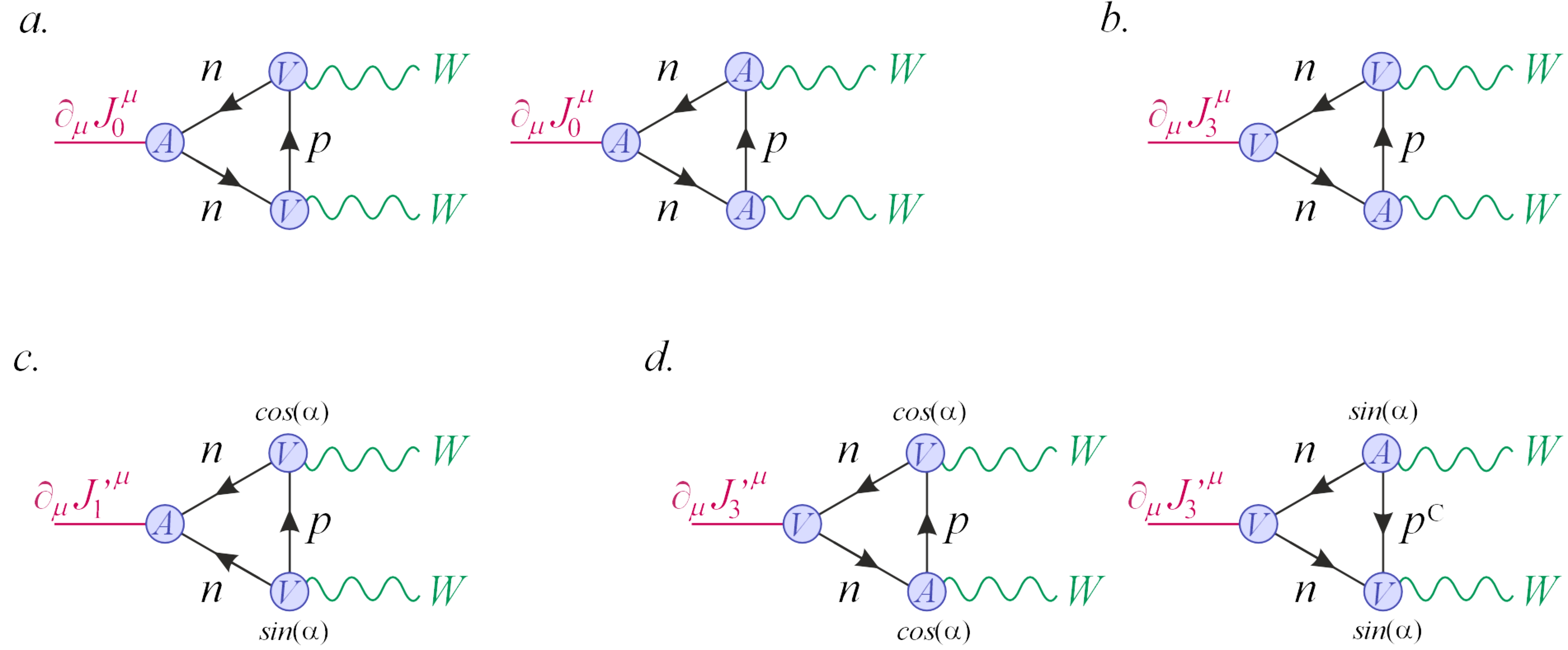}\caption{Top:
Anomalous triangle graphs contributing to the divergence of the axial current (a) and of the baryonic current (b). Bottom: In a scenario with only a Bogoliubov rotation $U(\alpha,0)$, the axial $\Delta\mathcal{B}=2$ current becomes anomalous because of the $\Delta\mathcal{B}=2$ components of the weak vertices (c), while the vector current anomaly gets reduced (d).}
\label{Fig1}
\end{figure}

In the absence of mass terms, the axial current gets an electroweak anomaly
from the $AVV$ and $AAA$ triangles (see Fig.~\ref{Fig1} and
Appendix~\ref{AppTriangles}), with the usual result
\end{subequations}
\begin{equation}
J_{0}^{\mu}=\bar{n}\gamma^{\mu}\gamma^{5}n\ ,\ \ \partial_{\mu}J_{0}^{\mu
}=g^{2}|V_{ud}|^{2}\frac{g_{V}^{2}+g_{A}^{2}}{16\pi^{2}}W^{+,\mu\nu}\tilde
{W}_{\mu\nu}^{-}\ .\label{dJ0}%
\end{equation}
The baryonic current is vectorial, but gets an anomalous divergence since the
weak gauge boson have both axial and vector couplings, see Fig.~\ref{Fig1}.
The triangle is again the $AVV$ one, but with gauge invariance now enforced on
the $A$ vertex and one of the $V$ vertices, giving%
\begin{equation}
J_{3}^{\mu}=-\bar{n}\gamma^{\mu}n\ ,\ \ \partial_{\mu}J_{3}^{\mu}%
=-g^{2}|V_{ud}|^{2}\frac{g_{A}g_{V}}{8\pi^{2}}W^{+,\mu\nu}\tilde{W}_{\mu\nu
}^{-}\ .\label{dJ3}%
\end{equation}
For both the axial and baryonic currents, the contributions from the proton
current should be added, but this is inessential here. For their part, the
$\Delta\mathcal{B}=2$ currents have no anomaly since the weak vertices are
$\Delta\mathcal{B}=0$, so
\begin{subequations}
\begin{align}
J_{1}^{\mu}  &  =(\bar{n}\gamma^{\mu}\gamma^{5}n^{\mathrm{C}}+\bar
{n}^{\mathrm{C}}\gamma^{\mu}\gamma^{5}n)\ ,\ \ \partial_{\mu}J_{1}^{\mu
}=0\ ,\\
J_{2}^{\mu}  &  =i(\bar{n}\gamma^{\mu}\gamma^{5}n^{\mathrm{C}}-\bar
{n}^{\mathrm{C}}\gamma^{\mu}\gamma^{5}n)\ ,\ \ \partial_{\mu}J_{2}^{\mu}=0\ .
\end{align}

Once turning on generic mass terms, these anomalies do not change, but do not
necessarily stay in the same current. Indeed, what is peculiar in this case is
that the anomaly moves around because the currents depend on the basis for the
neutron fields. Specifically, for neutron states in the standard basis, all
three currents are in general anomalous
\end{subequations}
\begin{equation}
\partial_{\mu}J_{i}^{\prime\mu}=\mathcal{R}(U)_{ij}\partial_{\mu}J_{j}^{\mu
}=\mathcal{R}(U)_{i3}\partial_{\mu}J_{3}^{\mu}\ ,\label{RJano}%
\end{equation}
with $\partial_{\mu}J_{3}^{\mu}$ given in Eq.~(\ref{dJ3}). Of course, this is
an artifact of the basis, but it must nevertheless be accounted for. Indeed,
though phenomenologically, it is always possible to reach the standard basis
via some $U$ transformation, the anomalous divergences do depend on the path
taken from the initial Lagrangian down to that basis. In this way, some
dependences on the original basis are still present.

Physically, this dependence boils down to that of the weak vertices. They also
keep a trace of the $U$ transformation, see Eq.~(\ref{Fullnnc}). The two
crucial points here is that they can develop a $\Delta\mathcal{B}=2$
component, and that the Bogoliubov rotation involves a $\gamma_{5}$, see
Eq.~(\ref{BogoN}). Let us take an example. We consider just a single
$U=U(\alpha,0)$. In the new basis, both $J_{1}^{\mu}$ and $J_{3}^{\mu}$ are
anomalous. Looking at Fig.~\ref{Fig1} it is clear that
\begin{align}
\partial_{\mu}J_{1}^{\prime\mu}  &  =2\cos\alpha\sin\alpha\partial_{\mu}%
J_{3}^{\mu}\ ,\\
\partial_{\mu}J_{3}^{\prime\mu}  &  =(\cos^{2}\alpha-\sin^{2}\alpha
)\partial_{\mu}J_{3}^{\mu}\ .
\end{align}
At the same time, $\mathcal{R}(U(\alpha,0))$ is a rotation matrix of angle
$2\alpha$, see Eq.~(\ref{RU}), so Eq.~(\ref{RJano}) gives the same result,
$\partial_{\mu}J_{1}^{\prime\mu}=\sin2\alpha\partial_{\mu}J_{3}^{\mu}$ and
$\partial_{\mu}J_{3}^{\prime\mu}=\cos2\alpha\partial_{\mu}J_{3}^{\mu}$. Yet,
it is quite satisfying to see that these trigonometric functions of $2\alpha$,
coming from the representation of the $SU(2)$ matrix $U(\alpha,0)$ as an
$SO(3)$ rotation matrix, precisely match the various possibilities to
construct the triangle graphs, with $\cos\alpha$ or $\sin\alpha$ at the weak
vertices, see Fig.~\ref{Fig1}.

\subsubsection{Anomalous transformations}

In full generality, the SM Lagrangian should include the CP-violating
couplings%
\begin{equation}
\mathcal{L}_{\theta}=\frac{g}{4\pi^{2}}\theta_{L}W^{\mu\nu}\tilde{W}_{\mu\nu
}+\frac{g^{\prime}}{4\pi^{2}}\theta_{Y}B^{\mu\nu}\tilde{B}_{\mu\nu}\ .
\end{equation}
Usually, the $W^{\mu\nu}\tilde{W}_{\mu\nu}$ term is cancelled by performing a
$\mathcal{B}+\mathcal{L}$ rotation, which is a symmetry of the SM, leaving the
innocuous abelian $B^{\mu\nu}\tilde{B}_{\mu\nu}$ term. This cannot be done
here since $\mathcal{B}$ is explicitly broken, and so is presumably
$\mathcal{L}$ to induce neutrino masses. Actually, if both these breaking are
induced spontaneously, as in Ref.~\cite{Arias-Aragon:2022byr}, then one could
imagine $W^{\mu\nu}\tilde{W}_{\mu\nu}$ is initially absent at the high scale,
but gets entirely generated by the diagonalization of neutron and neutrino
masses. Indeed, anomalous rephasing permits to move back-and-forth the
CP-violating phases in the mass terms and in the anomalous couplings. This is
the exact weak-interaction analog of how the strong $\theta$ parameter tuning
the gluonic $G^{\mu\nu}\tilde{G}_{\mu\nu}$ term is entangled with the phases
of the quark masses.

Our goal here is to calculate the induced $W^{\mu\nu}\tilde{W}_{\mu\nu}$
coupling arising from the diagonalization of the neutron mass matrix, i.e., by
the matrix $U$ of Eq.~(\ref{TotU}), which we can write as%
\begin{equation}
U=e^{i(\beta+\phi_{\Delta})\sigma_{3}/2}\cdot e^{i\alpha\sigma_{2}}\cdot
e^{-i\eta\sigma_{3}/2}\cdot e^{i(\pi/4)\sigma_{2}}\cdot e^{i\gamma\sigma
_{3}/2}\cdot e^{-i(\pi/4)\sigma_{2}}\cdot e^{i\zeta\sigma_{0}/2}\ .
\end{equation}
The chiral phase commutes with all the others and can easily be treated: it
induces the change in the Lagrangian%
\begin{align}
\delta_{\mathcal{C}}\mathcal{L}  &  =\zeta\partial_{\mu}J_{0}^{\mu}%
=g^{2}|V_{ud}|^{2}\frac{g_{V}^{2}+g_{A}^{2}}{16\pi^{2}}\zeta W^{+,\mu\nu
}\tilde{W}_{\mu\nu}^{-}\nonumber\\
&  =g^{2}|V_{ud}|^{2}\frac{g_{V}^{2}+g_{A}^{2}}{16\pi^{2}}W^{+,\mu\nu}%
\tilde{W}_{\mu\nu}^{-}\frac{m_{L}m_{R}\sin2\phi_{\Sigma}}{2m_{D}^{2}%
}+\mathcal{O}(m_{D}^{-3})\ .\label{dLchiral}%
\end{align}
As said before, we consider that $\zeta$ is induced only by Majorana phases.
The contributions coming from the phase of the Dirac mass terms for both the
neutron (denoted $\phi_{D}$ in Eq.~(\ref{mDexp})) and proton would simply add
up to $\zeta$, but are assumed to be absent here, as is consistent with the
splitting between their EDM and MDM.

We would like to proceed similarly for the other factors of $U$, but the
presence of Bogoliubov rotations complicates matters. Indeed, performing a
$P_{\mathcal{B}}(\phi)$ rephasing after such a rotation does not induces a
phase difference of $\phi$ between $n$ and $n^{\mathrm{C}}$ since these
rotated neutron states are mixtures of both $n$ and $n^{\mathrm{C}}$, see
Eq.~(\ref{BogoN}). After a Bogoliubov rotation, $P_{\mathcal{B}}(\phi)$ is no
longer purely baryonic but contains also a non-anomalous trivial global
rephasing component.

To view this explicitly, consider the transformation $U(\alpha,\sigma
)=P_{\mathcal{B}}(-\sigma)\cdot U(\alpha,0)\cdot P_{\mathcal{B}}(\sigma)$. If
$U(\alpha,0)$ had no impact, the baryonic rephasing $P_{\mathcal{B}}(-\sigma)$
would be compensated by $P_{\mathcal{B}}(\sigma)$, and no shift would occur
for the $W^{+,\mu\nu}\tilde{W}_{\mu\nu}^{-}$ coupling. This cannot hold
because if $\alpha=\pi/2$, we can write%
\begin{equation}
U(\pi/2,\sigma)=P_{\mathcal{B}}(-\sigma)\cdot U(\pi/2,0)\cdot P_{\mathcal{B}%
}(\sigma)=U(\pi/2,0)\cdot P_{\mathcal{B}}(2\sigma)=P_{\mathcal{B}}%
(-2\sigma)\cdot U(\pi/2,0)\ .
\end{equation}
The last expression shows that the true shift in $\theta_{L}$ musts actually
be $-2\sigma$. Indeed, $P_{\mathcal{B}}(-2\sigma)$ occurs in the original weak
interaction basis, and is followed by a non-anomalous real rotation
$U(\pi/2,0)$. That it is not anomalous is ensured by the expression%
\begin{equation}
U(\alpha,\sigma)=\cos\alpha\sigma_{0}-i\sigma_{1}(\sin\alpha\sin
\sigma)+i\sigma_{2}\sin\alpha\cos\sigma\ .
\end{equation}
Thus, $U(\pi/2,0)=\exp(i\pi\sigma_{2}/2)$ is associated to the current
$J_{2}^{\mu}$, which has no anomaly since we are still in the weak basis after
the $P_{\mathcal{B}}(-2\sigma)$ rotation (provided the proton is also rephased accordingly).

Returning to $U(\alpha,\sigma)$, it has an anomaly because it does not
exponentiate to $\exp(i\sigma_{1}a_{1})\cdot\exp(i\sigma_{2}a_{2})$ for some
$a_{1}$ and $a_{2}$. Instead, contributions proportional to $[\sigma
_{1},\sigma_{2}]=2i\sigma_{3}$ appear, and $\sigma_{3}$ corresponds to the
vector current which does have an anomaly in the weak basis, see
Eq.~(\ref{RJano}). Explicitly, with the decomposition $U(\alpha,\sigma
)=\exp(-i\sigma\sigma_{3}/2)\cdot\exp(i\alpha\sigma_{2})\cdot\exp
(i\sigma\sigma_{3}/2))$, the first factor generates $-\sigma\partial_{\mu
}J_{3}^{\mu}$ with $\partial_{\mu}J_{3}^{\mu}$ in the weak basis,
$U(\alpha,0)$ generates $\alpha\partial_{\mu}J_{2}^{\mu}$ which is not
anomalous, and the last factor generates $+\sigma\partial_{\mu}J_{3}%
^{\prime\mu}$ in the rotated basis, such that $\partial_{\mu}J_{3}^{\prime\mu
}=\mathcal{R}(U(\alpha,0))_{3i}\partial_{\mu}J_{i}^{\mu}$ with $\mathcal{R}%
(U(\alpha,0))$ the next-to-last rotation matrix in Eq.~(\ref{RU}). Since only
$\partial_{\mu}J_{3}^{\mu}\neq0$, we only need $\mathcal{R}(U(\alpha
,0))_{33}=\cos(2\alpha)$ to find%
\begin{equation}
\delta_{U(\alpha,\sigma)}\mathcal{L}=g^{2}|V_{ud}|^{2}\frac{g_{A}g_{V}}%
{8\pi^{2}}\left(  -\sigma+\sigma\cos(2\alpha)\right)  =g^{2}|V_{ud}|^{2}%
\frac{g_{A}g_{V}}{8\pi^{2}}\left(  -2\sigma\sin^{2}\alpha\right)  \ .
\end{equation}
Thus, $\delta_{U(\alpha,\sigma)}\mathcal{L}$ vanishes for $\alpha=0$, and is
proportional to $-2\sigma$ for $\alpha=\pi/2$, as it should.

A similar strategy applies for a generic change of basis. Specifically, we can
write%
\begin{equation}
U=\prod_{k=1}^{n}e^{i\theta_{k}\sigma_{i_{k}}}:\delta\mathcal{L}_{\mathcal{B}%
}=\sum_{k=1}^{n}\theta_{k}\partial_{\mu}J_{i_{k}}^{(k-1)\mu}\ ,\ \ J_{i}%
^{(k)\mu}=\mathcal{R}_{ij}(\prod_{l=1}^{k}e^{i\theta_{l}\sigma_{i_{l}}})\cdot
J_{j}^{\mu}\ ,\label{Noether}%
\end{equation}
where $\mathcal{R}(U)$ is constructed as in Eq.~(\ref{RU}). With the anomalous
term coming entirely from $\partial_{\mu}J_{3}^{\mu}$, we find
\begin{equation}
\delta\mathcal{L}_{\mathcal{B}}=\frac{g_{A}g_{V}}{8\pi^{2}}W^{+,\mu\nu}%
\tilde{W}_{\mu\nu}^{-}\sum_{k=1}^{n}\theta_{k}\mathcal{R}_{i_{k}3}(\prod
_{l=1}^{k-1}e^{i\theta_{l}\sigma_{i_{l}}})\ .
\end{equation}

An important caveat though is the use of Noether's theorem, which assumes the
field variation is small enough to discard higher-order corrections. The
$\pi/4$ rotations in Eq.~(\ref{RU}) are particularly problematic, so we write
instead
\begin{equation}
U=e^{i(\beta+\phi_{\Delta})\sigma_{3}/2}\cdot e^{i\alpha\sigma_{2}}\cdot
e^{-i\eta\sigma_{3}/2}\cdot e^{-i\gamma\sigma_{1}/2}\cdot e^{i\zeta\sigma
_{0}/2}\ .
\end{equation}
This form and Eq.~(\ref{RU}) do not give the same result, which can be
understood intuitively by the curvature of the sphere on which these rotations
live. That curvature cannot be caught by the linear approximation in
Eq.~(\ref{Noether}). In this respect, the above form is certainly more
appropriate since $\gamma$ is small, of $\mathcal{O}(m_{D}^{-3})$, see
Eq.~(\ref{mDexp}). The result is then%
\begin{align}
\delta\mathcal{L}_{\mathcal{B}}  &  =\frac{g_{A}g_{V}}{8\pi^{2}}W^{+,\mu\nu
}\tilde{W}_{\mu\nu}^{-}\left(  \beta+\phi_{\Delta}-\eta\cos2\alpha-\gamma
\sin2\alpha\cos\eta\right) \nonumber\\
&  =\frac{g_{A}g_{V}}{8\pi^{2}}W^{+,\mu\nu}\tilde{W}_{\mu\nu}^{-}\left(
\beta+\phi_{\Delta}-\eta\right)  +\mathcal{O}(m_{D}^{-2})\ .
\end{align}

We can compare this to the result one would get going back to the full
expression for $U$. If we extract the total baryonic phase as the phase
difference between $n$ and $n^{\mathrm{C}}$ induced by the full
transformation, but neglect the small admixture of the complex conjugated
states, we get from Eq.~(\ref{Fullnnc})%
\begin{align}
2\arg U_{22}^{\ast}  &  =2\arctan\frac{\cos\alpha\cos\gamma/2\sin(\beta
+\phi_{\Delta}-\eta)/2-\sin\alpha\sin\gamma/2\cos(\beta+\phi_{\Delta}+\eta
)/2}{\sin\alpha\sin\gamma/2\sin(\beta+\phi_{\Delta}+\eta)/2+\cos\alpha
\cos\gamma/2\cos(\beta+\phi_{\Delta}-\eta)/2}\nonumber\\
&  =\beta+\phi_{\Delta}-\eta-\gamma\tan\alpha\cos\eta+\mathcal{O}(\gamma
^{2})\ .
\end{align}
This result matches the previous one at the linear order only. Yet, the fact
that the dependence on $\gamma$ disappears when $\alpha=0$ is well-reproduced
in both representations (and would not be present had we used Eq.~(\ref{RU})).
This is appropriate since without the Bogoliubov rotation by $\alpha$, the
$\gamma$ rephasing has no anomaly. It essentially takes place in the weak
basis, and rephases identically both $n$ and $n^{\mathrm{C}}$. Since $\alpha$
is $\mathcal{O}(m_{D}^{-1})$, and $\gamma$ is $\mathcal{O}(m_{D}^{-3})$, this
phase is in any case totally negligible. So, using Eq.~(\ref{bmd}), our final
result is%
\begin{equation}
\delta\mathcal{L}_{\mathcal{B}}=\frac{g_{A}g_{V}}{8\pi^{2}}W^{+,\mu\nu}%
\tilde{W}_{\mu\nu}^{-}\left(  \phi_{\Delta}+\arctan\left(  \frac{m_{R}-m_{L}%
}{m_{R}+m_{L}}\tan\phi_{\Sigma}\right)  \right)  +\mathcal{O}(m_{D}%
^{-2})\ .\label{dLbaryon}%
\end{equation}
This is much larger than the impact of a chiral phase in the Majorana mass
terms, see Eq.~(\ref{dLchiral}). Notice that if the $\arg U_{22}^{\ast}$
baryonic rephasing is also applied to protons, thereby maintaining the reality
of the $\Delta\mathcal{B}=0$ part of the weak vertices, $\delta\mathcal{L}%
_{\mathcal{B}}$ gets multiplied by two.

\section{Axion-induced $n-\bar{n}$ oscillations}

All the ingredients are now in place to return to the baryonic axion model.
Bringing the mass term into the standard basis, the axion Lagrangian of
Eq.~(\ref{Laxion0}) becomes%
\begin{align}
\mathcal{L}_{eff} &  =\frac{1}{2}\partial_{\mu}a\partial^{\mu}a+\bar{n}(i%
\!\not\!\partial-m)n-\frac{\varepsilon}{2}(\bar{n}^{\mathrm{C}}n+\bar{n}n^{\mathrm{C}%
})-\frac{\mu}{2}\bar{n}\sigma^{\mu\nu}nF_{\mu\nu}\nonumber\\
&  \ \ \ \ +\frac{\partial_{\mu}a}{2v}J_{i}^{\mu}\mathcal{R}(U)_{i3}%
+\mathcal{L}_{Jac}+\mathcal{L}_{weak}+\delta_{\mathcal{C}}\mathcal{L}%
+\delta_{\mathcal{B}}\mathcal{L}\ .
\label{FinalLagrangian}
\end{align}
The terms in the first line form the standard $n-\bar{n}$ oscillation
Lagrangian. We assume to be in the basis in which $m_{D}$ is real and $\mu$ is
the usual magnetic dipole moment, and neglect the small induced electric
dipole moment, see Eq.~(\ref{EDMMDM}). Then, as stated in
Ref.~\cite{Berezhiani:2015uya}, none of these terms depend on the precise form
of the initial mass term, since no reference to the diagonalization matrix remain.

On the contrary, all the terms in the second line except $\mathcal{L}_{Jac}$
depend on the specific form of the initial mass terms. The matrix
$\mathcal{R}(U)$, given in Eq.~(\ref{RotR}), is the induced mixing starting
from the baryonic vector current. Explicitly, the three currents are%
\begin{equation}
J_{1}^{\mu}=(\bar{n}\gamma^{\mu}\gamma^{5}n^{\mathrm{C}}+\bar{n}^{\mathrm{C}%
}\gamma^{\mu}\gamma^{5}n)\ ,\ \ J_{2}^{\mu}=i(\bar{n}\gamma^{\mu}\gamma
^{5}n^{\mathrm{C}}-\bar{n}^{\mathrm{C}}\gamma^{\mu}\gamma^{5}n)\ ,\ \ J_{3}%
^{\mu}=-\bar{n}\gamma^{\mu}n\ ,
\end{equation}
where $n$ and $n^{\mathrm{C}}$ are understood to be the redefined neutron
fields, and their coefficients are%
\begin{equation}
\mathcal{R}(U)_{i3}=\left(  \dfrac{m_{L}^{2}-m_{R}^{2}}{4m_{D}\varepsilon_{s}%
},\ -\dfrac{m_{R}m_{L}\sin2\phi_{\Sigma}}{2m_{D}\varepsilon_{s}},\ 1\right)
+\mathcal{O}(m_{D}^{-2})\ .\label{RUfin}%
\end{equation}
Notice that in the CP conserving limit, $\phi_{L,R}=0$ and $\mathcal{R}%
(U)_{23}=0$. In that limit, we further have $\mathcal{R}(U)_{13}=0$ if
$m_{L}=m_{R}$ since $U$ becomes trivial.

The weak interaction $\mathcal{L}_{weak}$ are also to be given in the standard
basis, so they include $\Delta\mathcal{B}=2$ pieces, see Eq.~(\ref{Fullnnc}).
The anomalous terms are first the Jacobian terms $\mathcal{L}_{Jac}$ given in
Eq.~(\ref{LjacB}), which involve the axion, and then the shifted $\theta$
terms $\delta_{\mathcal{C}}\mathcal{L}$ and $\delta_{\mathcal{B}}\mathcal{L}$
given in Eqs.~(\ref{dLchiral}) and~(\ref{dLbaryon}), respectively. Note that
all three currents are anomalous. As explained previously, if the
$a\rightarrow WW$ process is calculated, the anomalies in the triangle graphs
involving these three currents then cancel out precisely with the terms in
$\mathcal{L}_{Jac}$.

If only the terms in the first line above are considered, the usual
Schrodinger-like formalism for $n-\bar{n}$ oscillations is derived.
Specifically, the EoM for $n$ and $n^{\mathrm{C}}$ are
\begin{equation}
\left\{
\begin{array}
[c]{c}%
\left(  i\gamma^{\mu}\partial_{\mu}-m-\dfrac{\mu}{2}\sigma^{\mu\nu}F_{\mu\nu
}\right)  n=\varepsilon n^{\mathrm{C}}\ ,\\
\left(  i\gamma^{\mu}\partial_{\mu}-m+\dfrac{\mu}{2}\sigma^{\mu\nu}F_{\mu\nu
}\right)  n^{\mathrm{C}}=\varepsilon n\ .
\end{array}
\right.
\end{equation}
We are free to move to the Dirac representation, where $\gamma^{0}$ is
diagonal (see Eq.~(\ref{DiracRep})) and allows the construction of the
non-relativistic limit by pulling out the time variation of the fields.
Setting $i\partial_{\mu}=(i\partial_{t},-\mathbf{p})$, this takes the form%
\begin{equation}
i\frac{\partial}{\partial t}\left(
\begin{array}
[c]{c}%
n\\
n^{\mathrm{C}}%
\end{array}
\right)  =\left(
\begin{array}
[c]{cc}%
\gamma^{0}\left(  \boldsymbol{\gamma}\cdot\mathbf{p}+m+\dfrac{\mu}{2}%
\sigma^{\mu\nu}F_{\mu\nu}\right)  & \varepsilon\gamma^{0}\\
\varepsilon\gamma^{0} & \gamma^{0}\left(  \boldsymbol{\gamma}\cdot
\mathbf{p}+m-\dfrac{\mu}{2}\sigma^{\mu\nu}F_{\mu\nu}\right)
\end{array}
\right)  \left(
\begin{array}
[c]{c}%
n\\
n^{\mathrm{C}}%
\end{array}
\right)  \ .
\end{equation}

A generic non-relativistic expansion corresponds to a unitary transformation
$n\rightarrow e^{iS}n$ and $n^{\mathrm{C}}\rightarrow e^{iS^{\prime}%
}n^{\mathrm{C}}$, such that the coupled equation become%
\begin{equation}
i\frac{d}{dt}\left(
\begin{array}
[c]{c}%
n\\
n^{\mathrm{C}}%
\end{array}
\right)  =\left(
\begin{array}
[c]{cc}%
e^{iS}\mathcal{H}e^{-iS}-\dot{S} & e^{iS}he^{-iS^{\prime}}\\
e^{iS^{\prime}}he^{-iS} & e^{iS^{\prime}}\mathcal{H}^{\mathrm{C}%
}e^{-iS^{\prime}}-\dot{S}^{\prime}%
\end{array}
\right)  \cdot\left(
\begin{array}
[c]{c}%
n\\
n^{\mathrm{C}}%
\end{array}
\right)  \ .
\end{equation}
To leading order, given the form of $\mathcal{H}$ and $\mathcal{H}%
^{\mathrm{C}}$, the transformations are identical, $iS=iS^{\prime
}=\boldsymbol{\gamma}\cdot\mathbf{p}/(2m)$, so%
\begin{align}
i\frac{d}{dt}\left(
\begin{array}
[c]{c}%
n\\
n^{\mathrm{C}}%
\end{array}
\right)   &  =\left(
\begin{array}
[c]{cc}%
\mathcal{H}+[iS,\mathcal{H}]+...-\dot{S} & h+[iS,h]+...\\
h+[iS,h]+... & \mathcal{H}^{\mathrm{C}}+[iS,\mathcal{H}^{\mathrm{C}}%
]+...-\dot{S}%
\end{array}
\right)  \cdot\left(
\begin{array}
[c]{c}%
n\\
n^{\mathrm{C}}%
\end{array}
\right) \nonumber\\
&  =\left(
\begin{array}
[c]{cc}%
\gamma^{0}\left(  m-\mu\boldsymbol{\sigma}\cdot\mathbf{B}\right)  &
\varepsilon\gamma^{0}\\
\varepsilon\gamma^{0} & \gamma^{0}\left(  m+\mu\boldsymbol{\sigma}%
\cdot\mathbf{B}\right)
\end{array}
\right)  \cdot\left(
\begin{array}
[c]{c}%
n\\
n^{\mathrm{C}}%
\end{array}
\right)  +\mathcal{O}(\mathbf{p}^{2}/m)\ ,\label{Schro1}%
\end{align}
since $\varepsilon$ is constant in space. A crucial property of the above
system is the diagonal nature of the $\Delta\mathcal{B}=2$ coupling
$\varepsilon\gamma^{0}$. If the magnetic field is aligned with the spin
quantization axis, this allows us to split the set of eight equations into
four equivalent sets of two equations. Each of these sets describe the
evolution of a two-state system, from which one recovers $P_{n\rightarrow
\bar{n}}(t)$ as given in Eq.~(\ref{Intro2}).

To the above picture we now add the axion derivative interactions. The vector
current does not lead to any effect because its whole impact can be encoded as
$(E,p)\rightarrow(E-\dot{a}/m,p-\mathbf{\nabla}a/m)$. This reflects the fact
that such a current corresponds to the pure gauge part of the electromagnetic
minimal coupling for a charged fermion. Even if the neutron is neutral, such a
would-be gauge part cancels out, and none of these terms can lead to
oscillations. More problematic are the $\Delta\mathcal{B}=2$ currents because
of the presence of the $\gamma^{5}$. Specifically, consider the $J_{1}^{\mu}$
current, which adds to the equation of motion the terms:%
\begin{equation}
\left\{
\begin{array}
[c]{c}%
\left(  i\gamma^{\mu}\partial_{\mu}-m-\dfrac{\mu}{2}\sigma^{\mu\nu}F_{\mu\nu
}\right)  n=\varepsilon n^{\mathrm{C}}+\dfrac{\partial_{\mu}a}{4v}%
\lambda\gamma^{\mu}\gamma^{5}n^{\mathrm{C}}\ ,\\
\left(  i\gamma^{\mu}\partial_{\mu}-m+\dfrac{\mu}{2}\sigma^{\mu\nu}F_{\mu\nu
}\right)  n^{\mathrm{C}}=\varepsilon n+\dfrac{\partial_{\mu}a}{4v}%
\lambda\gamma^{\mu}\gamma^{5}n\ ,
\end{array}
\right.
\end{equation}
where $\lambda=\mathcal{R}(U)_{13}$ of Eq.~(\ref{RUfin}). The off-diagonal
entries in the Schrodinger equation~(\ref{Schro1}) are then
\begin{equation}
\varepsilon\gamma^{0}\rightarrow\varepsilon\gamma^{0}+\frac{i\lambda}%
{4v}(\partial_{t}a\gamma^{5}+\boldsymbol{\sigma}\cdot\mathbf{\nabla}a)\ .
\end{equation}
This time, diagonalizing the $\Delta\mathcal{B}=0$ sector does not diagonalize
the $\Delta\mathcal{B}=2$ sector. As a result, the set of eight equations no
longer splits into equivalent subsets of two equations. The $\partial_{t}a$
term mixes small and large components: it couples e.g. the large spinor
component of $n$ to the small spinor component of $n^{\mathrm{C}}$. How to
proceed in such a case probably requires dealing with the whole set of
equations, and the usual simple formalism for $n-\bar{n}$ oscillations is not
sufficient. This goes beyond our purpose, so instead, we simply notice that
had we used the elimination method, that small component would appear
suppressed by $\boldsymbol{\sigma}\cdot\mathbf{p}/m$, and thus working at
leading order, it can be discarded.

The axion wind term, proportional to $\mathbf{\nabla}a$, is immediately
diagonal, but not proportional to $\gamma_{0}$. It affects differently
neutrons and antineutrons, contrary to $\varepsilon$, and depends on the
neutron polarization state through $\boldsymbol{\sigma}$. Let us thus consider
the neutrons, and in addition, turn off the parameter $\varepsilon$. This can
be achieved by setting $\phi_{\Sigma}=\pi/2$ and $m_{L}=m_{R}=\bar{m}$. This
forces $\varepsilon=0$ exactly, while $J_{1}^{\mu}$ is tuned by $\lambda
=\mathcal{R}(U)_{13}=\bar{m}/m_{D}$. This scenario is not necessarily
violating CP since $\phi_{\Delta}$ is still free. If we set it also to $\pi
/2$, $\phi_{L}=0$ but $\phi_{R}=\pi$, so $m_{R}$ (or $m_{L}$) simply ends up
with a "wrong sign". There is no $J_{2}^{\mu}$ component for $\phi_{\Sigma
}=\pi/2$, $\mathcal{R}(U)_{23}=0$, as expected since the original baryonic
current is CP conserving. With all this, the oscillation equation reduces to%
\begin{equation}
i\frac{d}{dt}\left(
\begin{array}
[c]{c}%
n\\
n^{\mathrm{C}}%
\end{array}
\right)  =\left(
\begin{array}
[c]{cc}%
\gamma^{0}\left(  m-\mu\boldsymbol{\sigma}\cdot\mathbf{B}\right)  &
\dfrac{i\lambda}{4v}\boldsymbol{\sigma}\cdot\mathbf{\nabla}a\\
\dfrac{i\lambda}{4v}\boldsymbol{\sigma}\cdot\mathbf{\nabla}a & \gamma
^{0}\left(  m+\mu\boldsymbol{\sigma}\cdot\mathbf{B}\right)
\end{array}
\right)  \cdot\left(
\begin{array}
[c]{c}%
n\\
n^{\mathrm{C}}%
\end{array}
\right)  +\mathcal{O}(\mathbf{p}/m)\ .
\end{equation}
This scenario collapses to the resonant case discussed in the Introduction,
Yet, instead of having $\varepsilon=\varepsilon(t)$, what matters is the time
dependence of the gradient of the axion field, coupled to the neutron spin, so
polarized neutrons would be needed.

Numerically, the presence of the gradient is a high price to pay. To compare
with Eq.~(\ref{Intro4}), let us write the classical dark matter axion field is
$a(x,t)\approx a_{0}\sin(m_{a}(\mathbf{v}_{a}\cdot\mathbf{x}+t))$ with
$v_{a}\sim10^{-3}$ the local axion dark matter speed relative to the sun, and
$a_{0}=\sqrt{2\rho_{DM}}/m_{a}$ with $\rho_{DM}=0.4\ $GeV$/cm^{3}%
~$\cite{Catena:2009mf}. Neglecting the orientation of the
Earth~\cite{Knirck:2018knd} and neutron spins, we can identify%
\begin{equation}
\varepsilon_{0}\sin(\omega t)\approx\dfrac{\lambda}{4v}m_{a}v_{a}a_{0}%
\sin(m_{a}t)\ .
\end{equation}
The specific axion model enters in the parameter $\lambda=\mathcal{R}%
(U)_{13}=\bar{m}/m_{n}$. In the simplest case of Eq.~(\ref{Bmod1}), Majorana
masses are given in terms of the PQ breaking scale as $\bar{m}=m_{L}%
=m_{R}\equiv\bar{\varepsilon}v$. Then, the size of the off-diagonal entries
become independent of the axion parameters:%
\begin{equation}
\varepsilon_{0}\approx\frac{v_{a}\sqrt{2\rho_{DM}}}{m_{n}}\bar{\varepsilon
}\approx(10^{-15}\times\bar{\varepsilon})~\text{eV\ }.
\end{equation}
This cannot lead to observational effects because even in a scenario in which
there is no oscillation in vacuum, $\lambda=\bar{\varepsilon}v/m_{n}\ll1$, so
$\bar{\varepsilon}$ must be tiny for $v\approx10^{12}$~GeV. Indeed, as
discussed in Sec.~\ref{weak}, $n-\bar{n}$ mixing in production and decay is
maximal when oscillations in vacuum are turned off, with a probability scaling
like $\lambda^{2}$. Naively, the ILL
search~\cite{Baldo-Ceolin:1989vpk,Baldo-Ceolin:1994hzw} sets $\lambda<10^{-9}$
so $\bar{\varepsilon}<10^{-21}$. Returning to $n-\bar{n}$ oscillations, this
would give $\varepsilon_{0}<10^{-36}$, way too small to be observable. Given
that $\varepsilon_{0}$ is rather insensitive to axion parameters, this
conclusion appears robust.

True QCD axion cannot impact $n-\bar{n}$ oscillations, but the situation would
evidently be totally different for axion-like particles (ALP). If a
pseudoscalar directly couples to the $\Delta\mathcal{B}=2$ axial currents, all
other $\Delta\mathcal{B}=2$ mixing effects would be absent, whether in the
weak vertices or in oscillations, and $\lambda/v$ would be a free parameter
not related to Majorana masses. For the QCD axion, this setting is problematic
because though the axion shift symmetry is correctly built in, there is no way
to construct an equivalent exponential parametrization (see Sec.~\ref{dbcurr}%
). The PQ symmetry would have to be broken in a quite convoluted way, and it
may even be impossible to do so without at some level inducing
axion-independent $\Delta\mathcal{B}=2$ couplings. This is why we consider
that turning on only the $\Delta\mathcal{B}=2$ derivative coupling is in
conflict with solving the strong CP puzzle. Yet, from a model-independent
perspective, this scenario is perfectly viable and worth the effort from an
experimental point of view.

\section{Conclusion}

In this paper, motivated by the possibility to induce resonant $n-\bar{n}$
oscillations if dark matter carries $\mathcal{B}=2$, we have performed a
detailed analysis of axion models in which the PQ symmetry is aligned with
baryon number. The Goldstone boson nature of the axion severely constrains
these theoretical constructions, to the point that it appears
phenomenologically impossible to generate sizeable effects in $n-\bar{n}$
oscillations. The reasons for this are however not trivial, and in the course
of this analysis, several delicate points were clarified:

\begin{itemize}
\item Though oscillations in vacuum are always parametrically the same no
matter the initial neutron Dirac and Majorana mass terms, imprints of the path
taken to bring these mass terms to the standard form are left in many sectors.
In particular, neutron production and decay, the baryon number current, and
triangle anomalies all depend on the rotation needed to reach the standard
$n-\bar{n}$ basis. This is often overlooked in the literature.

\item In the present context, these imprints play a central role, and are
embodied in our final Lagrangian, Eq.~(\ref{FinalLagrangian}). Indeed, true
axion models necessarily entangle the $\Delta\mathcal{B}=2$ mass terms with
the axion couplings. Axion induced oscillations are then always superseded
either by $n-\bar{n}$ oscillations in vacuum, or $n-\bar{n}$ mixing in weak
decays. Both are experimentally very constrained, and together, they cover the
whole parameter space.

\item Actually, we can draw a much stronger conclusion. The fact that
axion-induced oscillations are small could have been expected from the
shift-symmetry, bringing in a factor $m_{a}v_{a}/v$ with $v_{a}$ the axion
velocity, $m_{a}$ its mass, and $v$ the PQ breaking scale. But this is not the
full story. Even in the relativistic limit, at higher energies or in
astrophysical environments, the $\partial_{\mu}a\bar{n}^{\mathrm{C}}%
\gamma^{\mu}\gamma^{5}n$ coupling is tiny. It is not only suppressed by $v$,
as would be expected for dimensional reason, but in addition by a coupling
constant known experimentally to be tiny, $\mathcal{R}(U)_{13}$ and
$\mathcal{R}(U)_{23}<10^{-9}$ in Eq.~(\ref{FinalLagrangian}), because true
axion models require $\partial_{\mu}a\bar{n}^{\mathrm{C}}\gamma^{\mu}%
\gamma^{5}n$ to always be accompanied by axionless $n-\bar{n}$ mixing effects.

\item Another way to view this obstruction is as due to the very specific
relationship between the two couplings $a\bar{n}^{\mathrm{C}}\gamma^{5}n$ and
$\partial_{\mu}a\bar{n}^{\mathrm{C}}\gamma^{\mu}\gamma^{5}n$ of
Eq.~(\ref{IntroC2}). We investigated in details how the reparametrization
theorem works for $\Delta\mathcal{B}=2$ couplings, and found it to be far less
trivial than for the usual chiral axion models. Said differently, resonant
$n-\bar{n}$ oscillations can be induced via either $a\bar{n}^{\mathrm{C}%
}\gamma^{5}n$ or $\partial_{\mu}a\bar{n}^{\mathrm{C}}\gamma^{\mu}\gamma^{5}n$,
except if they share the reparametrization relationship expected for the QCD
axion. In this respect, the possibility thus remains open for an axion-like
particle, for which there would be no associated axionless $n-\bar{n}$ mass
mixing effects. In that context, one would presumably concentrate on the
shift-symmetric $\partial_{\mu}a\bar{n}^{\mathrm{C}}\gamma^{\mu}\gamma^{5}n$
coupling, but both would be possible from a model-independent perspective.
\end{itemize}

In our opinion, this settles the question for axion-induced $n-\bar{n}$ oscillations in most models. Maybe some intricate ways to break the PQ symmetry could evade this conclusion, but they remain to be devised. Beyond that, some questions remain open, both on the technical aspects or concerning other related signatures, that would deserve dedicated studies. Let us mention in particular:

\begin{itemize}
\item The formalism developed here would apply equally well to neutrinos. In
addition, it is quite natural to merge the Majoron and axion
mechanism~\cite{Clarke:2015bea,Quevillon:2020hmx}, thereby creating axion
couplings to right-handed neutrinos. In that context, reparametrization of the
neutrino would couple the axion to the lepton number current, and the
diagonalization of the neutrino mass would mix that current with
$\Delta\mathcal{L}=2$ axial currents. The main difference between the two
sectors is of course the expected scaling between Majorana and Dirac mass
terms, which are somehow opposite. These scenarios would be worth further investigations.

\item The usual non-relativistic reduction for the $n-\bar{n}$ system breaks
down for pseudoscalar couplings like $a\bar{n}^{\mathrm{C}}\gamma^{5}n$ and
$\partial_{\mu}a\bar{n}^{\mathrm{C}}\gamma^{\mu}\gamma^{5}n$. The problem
boils down to axial mixing terms between $n$ and $n^{\mathrm{C}}$, that
prevents the immediate elimination of their small spinor components, or the
trivial decoupling of their negative-energy components when using the
Foldy-Wouthuysen formalism. Though we think we correctly identified the
leading non-relativistic effect, further work would be needed to develop a
systematic procedure for this system.

\item Fundamentally, baryon number is a flavored $U(1)$ symmetry, and as such,
there is no reason to expect it to be universal in flavor space. Rather, as
advocated in Ref.~\cite{Nikolidakis:2007fc,Smith:2011rp,Durieux:2012gj}, it
would even make more sense to expect a coupling to flavor singlet combinations
of quarks, like e.g. $dst\times dst$ instead of $ddu\times ddu$. The latter
would then not be forbidden, but very suppressed by SM flavor mixings. This
means that $\lambda$ would indeed be tiny, so there would be no induced
$n-\bar{n}$ oscillations, but it would open the door to signatures involving
heavy quarks, that could be looked for at colliders or in heavy baryon
transitions~\cite{Alonso-Alvarez:2021oaj}.
\end{itemize}

To conclude, let us stress that scalar or pseudoscalar-induced $\Delta
\mathcal{B}=2$ resonant effects in $n-\bar{n}$ represents an unexplored
possible signature of dark matter. Though unifying such scenarios with axions
appear impossible at present, this does not mean Nature has not chosen that
path. Experimental searches for such a striking signature would really be
worth the effort.

\subsubsection*{Acknowledgements:}

First and foremost, we would like to thank Sacha Davidson for her help and
many advises in the early stage of this work. This research is supported by
the IN2P3 Master project \textquotedblleft Axions from Particle Physics to
Cosmology\textquotedblright, and from the French National Research Agency
(ANR) in the framework of the \textquotedblleft GrAHal\textquotedblright
project (ANR-22-CE31-0025).

\appendix

\section{Appendices}

\subsection{Conventions\label{appConv}}

Charge conjugate states are defined as
\begin{equation}
\psi^{\mathrm{C}}=\mathrm{C}\bar{\psi}^{T}\ ,\ \ \bar{\psi}^{\mathrm{C}}%
=-\psi^{T}\mathrm{C}^{\dagger}\ ,\label{DEF1}%
\end{equation}
with $\mathrm{C}$ satisfying $\mathrm{C}^{\dagger}=\mathrm{C}^{-1}%
,\;\mathrm{C}^{T}=-\mathrm{C}$. While under hermitian conjugation, bilinear
currents transform as $(\bar{\psi}_{2}\Gamma_{A}\psi_{1})^{\dagger}=\bar{\psi
}_{1}\Gamma_{A}\psi_{2}$, their behavior under charge conjugation involves
specific signs:%
\begin{equation}
(\bar{\psi}_{2}^{\mathrm{C}}\Gamma_{A}\psi_{1}^{\mathrm{C}})=\eta_{A}%
(\bar{\psi}_{1}\Gamma_{A}\psi_{2})\ ,\ \ \eta_{A}=+1\text{ for }\Gamma
_{A}=1,i\gamma_{5},\gamma^{\mu}\gamma_{5},\ \ \eta_{A}=-1\text{ for }%
\Gamma_{A}=\gamma^{\mu},\sigma^{\mu\nu}\ ,\label{DEF2}%
\end{equation}
since $\mathrm{C}^{-1}\Gamma_{A}\mathrm{C}=\eta_{A}\Gamma_{A}^{T}$. Chiral
projections get inverted under charge conjugation, so with $P_{L,R}%
=(1\mp\gamma_{5})/2$,%
\begin{equation}
n_{L,R}=P_{L,R}n,\ \ \bar{n}_{R,L}=\bar{n}P_{L,R}\Longrightarrow
n_{R,L}^{\mathrm{C}}=P_{L,R}n^{\mathrm{C}},\bar{n}_{L,R}^{\mathrm{C}}=\bar
{n}^{\mathrm{C}}P_{L,R}\ .\label{DEF3}%
\end{equation}

In the present work, we adopt the Weyl representation%
\begin{equation}
\gamma^{\mu}=\left(
\begin{array}
[c]{cc}%
0 & (\sigma^{\mu})_{\alpha\dot{\beta}}\\
(\bar{\sigma}^{\mu})^{\dot{\alpha}\beta} & 0
\end{array}
\right)  \ ,\ \ \sigma^{\mu\nu}=\left(
\begin{array}
[c]{cc}%
(\sigma^{\mu\nu})_{\alpha}^{\,\,\,\beta} & 0\\
0 & (\bar{\sigma}^{\mu\nu})_{\,\,\,\dot{\beta}}^{\dot{\alpha}}%
\end{array}
\right)  \ ,\ \ \gamma^{5}=\left(
\begin{array}
[c]{cc}%
-\mathbf{1} & 0\\
0 & \mathbf{1}%
\end{array}
\right)  \ ,
\end{equation}
with $\left(  \sigma^{\mu}\right)  _{\alpha\dot{\beta}}\equiv
(1,\boldsymbol{\sigma})$ and $\left(  \bar{\sigma}^{\mu}\right)  ^{\dot
{\alpha}\beta}\equiv(1,-\boldsymbol{\sigma})$. This permits to identify the
chiral components of a Dirac spinor with two-components Weyl spinors, see
Eq.~(\ref{WeylSpinor}). By contrast, the Dirac representation used to
construct the non-relativistic limit asks for $\gamma^{0}$ to be diagonal, in
which case
\begin{equation}
\gamma^{0}=\left(
\begin{array}
[c]{cc}%
\mathbf{1} & 0\\
0 & -\mathbf{1}%
\end{array}
\right)  \ ,\ \boldsymbol{\gamma}=\left(
\begin{array}
[c]{cc}%
0 & \boldsymbol{\sigma}\\
-\boldsymbol{\sigma} & 0
\end{array}
\right)  \ ,\ \ \gamma^{5}=\left(
\begin{array}
[c]{cc}%
0 & \mathbf{1}\\
\mathbf{1} & 0
\end{array}
\right)  \ .\label{DiracRep}%
\end{equation}

\subsection{Explicit form of the mixing matrix\label{ExplicitU}}

Explicitly, up to terms of $\mathcal{O}((m_{L,R}/m_{D})^{3})$, a relatively compact expression for $U$ is obtained as, for the chiral transformation:

\begin{align}
P_{\mathcal{C}}(\zeta)  &  =(-\phi_{D}+m_{L}m_{R}\sin2\phi_{\Sigma}%
/(2m_{D}^{2}))\mathbf{1}\ ,
\end{align}
and for the Bogoliubov and baryonic transformations:
\begin{align}
U(\alpha,-\beta)  &  =\left(
\begin{array}
[c]{cc}%
1-\dfrac{\varepsilon_{p}^{2}}{8m_{D}^{2}} & \dfrac{(m_{L}-m_{R})c_{\phi
_{\Sigma}}-i(m_{L}+m_{R})s_{\phi_{\Sigma}}}{4m_{D}}\\
\dfrac{(m_{R}-m_{L})c_{\phi_{\Sigma}}-i(m_{L}+m_{R})s_{\phi_{\Sigma}}}{4m_{D}}
& 1-\dfrac{\varepsilon_{p}^{2}}{8m_{D}^{2}}%
\end{array}
\right)  \ ,\\
P_{\mathcal{B}}(\delta-\eta)  &  =\left(
\begin{array}
[c]{cc}%
1+i\dfrac{m_{R}^{2}-m_{L}^{2}}{8\varepsilon_{s}^{2}}\dfrac{m_{L}m_{R}\sin
2\phi_{\Sigma}}{4m_{D}^{2}} & 0\\
0 & 1-i\dfrac{m_{R}^{2}-m_{L}^{2}}{8\varepsilon_{s}^{2}}\dfrac{m_{L}m_{R}%
\sin2\phi_{\Sigma}}{4m_{D}^{2}}%
\end{array}
\right)  \ ,\\
P_{\mathcal{B}}(\beta-\delta)  &  =\left(
\begin{array}
[c]{cc}%
\dfrac{\bar{\varepsilon}_{s}-i(m_{L}-m_{R})\sin\phi_{\Sigma}}{2\sqrt
{\bar{\varepsilon}_{s}\varepsilon_{s}}} & 0\\
0 & \dfrac{\bar{\varepsilon}_{s}+i(m_{L}-m_{R})\sin\phi_{\Sigma}}{2\sqrt
{\bar{\varepsilon}_{s}\varepsilon_{s}}}%
\end{array}
\right)  \ ,\
\end{align}
with $c_{\phi_{\Sigma}},s_{\phi_{\Sigma}}=\cos\phi_{\Sigma},\sin\phi_{\Sigma}
$ and $\bar{\varepsilon}_{s}=(m_{L}+m_{R})\cos\phi_{\Sigma}+2\varepsilon_{s}$.
These expressions are adequate to derive the series expansion when
$m_{L}\rightarrow m_{R}$, $\phi_{\Sigma}\rightarrow0$ or $m_{R}\rightarrow0$,
which can be written down straightforwardly. For $\phi_{\Sigma}$ small and
$\phi_{D}=0$, we find%
\begin{align}
U  &  =P_{\mathcal{B}}(\phi_{\Delta})\cdot\left[  \left(
\begin{array}
[c]{cc}%
1 & 0\\
0 & 1
\end{array}
\right)  -i\frac{\phi_{\Sigma}}{2}\frac{m_{L}-m_{R}}{m_{L}+m_{R}}\left(
\begin{array}
[c]{cc}%
1 & 0\\
0 & -1
\end{array}
\right)  +\frac{m_{L}-m_{R}}{4m_{D}}\left(
\begin{array}
[c]{cc}%
0 & 1\\
-1 & 0
\end{array}
\right)  \right. \nonumber\\
&  \ \ \ \ \ \ \ \ \ \ \ \ \ \ \ \ \ \ \ \ \ \ \ \ \ \left.  -i\frac
{\phi_{\Sigma}}{8}\frac{m_{L}^{2}+6m_{L}m_{R}+m_{R}^{2}}{(m_{L}+m_{R})m_{D}%
}\left(
\begin{array}
[c]{cc}%
0 & 1\\
1 & 0
\end{array}
\right)  +\mathcal{O}(m_{D}^{-2},\phi_{\Sigma}^{2})\right]  \ .
\end{align}
The expansion for $m_{L}\rightarrow m_{R}$ is very similar, as well as that
for $\phi_{\Sigma}\rightarrow\pi$. If one of the Majorana mass term is absent,
say $m_{R}\rightarrow0$, the $U$ matrix simplifies to%
\begin{equation}
U=\left(
\begin{array}
[c]{cc}%
e^{-i\phi_{L}/2} & 0\\
0 & e^{i\phi_{L}/2}%
\end{array}
\right)  +\frac{m_{L}}{4m_{D}}\left(
\begin{array}
[c]{cc}%
0 & e^{-i\phi_{L}/2}\\
-e^{i\phi_{L}/2} & 0
\end{array}
\right)  +\mathcal{O}(m_{D}^{-2},m_{R})\ .
\end{equation}
and analogously for $m_{L}\rightarrow0$.

\subsection{Triangle diagrams\label{AppTriangles}}

Most expressions in the literature deal with the triangle graph with one axial
current $A$, and two symmetrized vector current $V$. The usual discussion then
revolves around the issue of maintaining the vector symmetry, which fixes the
inherent ambiguity of the regularized triangle loop and leaves the whole
anomaly into the axial current $A$. This is insufficient for us since one of
the $V$ is to be the anomalous baryon current, while the other $V$ should be
the non-anomalous weak interaction. Similarly, the $AAA$ triangle is usually
described in its fully symmetrized form, while here two of the axial currents
have to be conserved.

General expressions for the $AVV$ and $AAA$ triangle graphs with axial and
vector currents were derived in Ref.~\cite{Quevillon:2019zrd}, following the
procedure detailed in Ref.~\cite{Weinberg:1996kr}. Those forms are independent
of which of the currents is assumed to be anomalous. For ease of reference,
those expressions are reproduced here. For the $AVV$ triangles, the general
Ward identities are
\begin{subequations}
\label{AnoAVVm}%
\begin{align}
i(q_{1}+q_{2})_{\alpha}\mathcal{T}_{AV_{1}V_{2}}^{\alpha\beta\gamma}  &
=2im\mathcal{T}_{PV_{1}V_{2}}^{\beta\gamma}+\frac{\operatorname*{Tr}%
(T_{A}\{T_{V_{1}},T_{V_{2}}\})}{8\pi^{2}}\left(  a-b\right)  \varepsilon
^{\beta\gamma\mu\nu}q_{1\mu}q_{2\nu}\ ,\\
-i(q_{1})_{\beta}\mathcal{T}_{AV_{1}V_{2}}^{\alpha\beta\gamma}  &
=\frac{\operatorname*{Tr}(T_{A}\{T_{V_{1}},T_{V_{2}}\})}{8\pi^{2}}\left(
1+b\right)  \varepsilon^{\gamma\alpha\mu\nu}q_{1\mu}q_{2\nu}\ ,\\
-i(q_{2})_{\gamma}\mathcal{T}_{AV_{1}V_{2}}^{\alpha\beta\gamma}  &
=\frac{\operatorname*{Tr}(T_{A}\{T_{V_{1}},T_{V_{2}}\})}{8\pi^{2}}\left(
1-a\right)  \varepsilon^{\alpha\beta\mu\nu}q_{1\mu}q_{2\nu}\ ,
\end{align}
and similarly, for the $AAA$ triangles:
\end{subequations}
\begin{subequations}
\label{AnoAAAm}%
\begin{align}
i(q_{1}+q_{2})_{\alpha}\mathcal{T}_{A_{1}A_{2}A_{3}}^{\alpha\beta\gamma}  &
=2im\mathcal{T}_{PA_{1}A_{2}}^{\beta\gamma}+\frac{\operatorname*{Tr}(T_{A_{1}%
}\{T_{A_{2}},T_{A_{3}}\})}{8\pi^{2}}\left(  a-b\right)  \varepsilon
^{\beta\gamma\mu\nu}q_{1\mu}q_{2\nu}\ ,\\
-i(q_{1})_{\beta}\mathcal{T}_{A_{1}A_{2}A_{3}}^{\alpha\beta\gamma}  &
=2im\mathcal{T}_{PA_{2}A_{3}}^{\alpha\gamma}+\frac{\operatorname*{Tr}%
(T_{A_{1}}\{T_{A_{2}},T_{A_{3}}\})}{8\pi^{2}}\left(  1+b\right)
\varepsilon^{\gamma\alpha\mu\nu}q_{1\mu}q_{2\nu}\ ,\\
-i(q_{2})_{\gamma}\mathcal{T}_{A_{1}A_{2}A_{3}}^{\alpha\beta\gamma}  &
=2im\mathcal{T}_{PA_{1}A_{3}}^{\alpha\beta}+\frac{\operatorname*{Tr}(T_{A_{1}%
}\{T_{A_{2}},T_{A_{3}}\})}{8\pi^{2}}\left(  1-a\right)  \varepsilon
^{\alpha\beta\mu\nu}q_{1\mu}q_{2\nu}\ ,
\end{align}
The parameters $a$ and $b$ are chosen to remove the anomalous piece from the
conserved currents. In these expressions, the mass-dependent pseudoscalar
triangle amplitudes are
\end{subequations}
\begin{subequations}
\label{TriPVVPAA}%
\begin{align}
\mathcal{T}_{PVV}^{\alpha\beta}  &  =-i\frac{\operatorname*{Tr}(T_{A}%
\{T_{V_{1}},T_{V_{2}}\})}{4\pi^{2}}mC_{0}(m^{2})\varepsilon^{\alpha\beta\mu
\nu}q_{1\mu}q_{2\nu}\ ,\\
\mathcal{T}_{PAA}^{\alpha\beta}  &  =-i\frac{\operatorname*{Tr}(T_{A_{1}%
}\{T_{A_{2}},T_{A_{3}}\})}{4\pi^{2}}m(C_{0}(m^{2})+2C_{1}(m^{2}))\varepsilon
^{\alpha\beta\mu\nu}q_{1\mu}q_{2\nu}\ ,
\end{align}
where $C_{i}(m^{2})=C_{i}(q_{1}^{2},q_{2}^{2},(q_{1}+q_{2})^{2},m^{2}%
,m^{2},m^{2})$. The operator versions of these amplitude-level identities is
obtained using $\langle\gamma(q_{1},\alpha_{1})\gamma(q_{2},\alpha_{2}%
)|F_{\mu\nu}\tilde{F}^{\mu\nu}|0\rangle=4ie^{\alpha_{1}\alpha_{2}\rho\sigma
}q_{1,\rho}q_{2,\sigma}$.

Adapted to our case, and with only a Dirac mass term, we find for the axial
current the two contributions:
\end{subequations}
\begin{align}
i(q_{1}+q_{2})_{\alpha}\mathcal{T}_{AVV}^{\alpha\beta\gamma}  &
=2im_{D}\mathcal{T}_{PVV}^{\beta\gamma}+g^{2}|V_{ud}|^{2}\frac{g_{V}^{2}%
}{16\pi^{2}}\varepsilon^{\beta\gamma\mu\nu}q_{1\mu}q_{2\nu}\ ,\\
i(q_{1}+q_{2})_{\alpha}\mathcal{T}_{A_{1}A_{2}A_{3}}^{\alpha\beta\gamma}  &
=2im_{D}\mathcal{T}_{PA_{1}A_{2}}^{\beta\gamma}+g^{2}|V_{ud}|^{2}\frac
{g_{A}^{2}}{16\pi^{2}}\varepsilon^{\beta\gamma\mu\nu}q_{1\mu}q_{2\nu}\ ,
\end{align}
and for the baryonic current the contribution (a factor two comes from the two
permutation of the axial and vector gauge vertex)%
\begin{equation}
-i(q_{1})_{\beta}\mathcal{T}_{AV_{1}V_{2}}^{\alpha\beta\gamma}=g^{2}%
|V_{ud}|^{2}\frac{g_{A}g_{V}}{8\pi^{2}}\varepsilon^{\gamma\alpha\mu\nu}%
q_{1\mu}q_{2\nu}\ .
\end{equation}
From this, the Ward identities in Eqs.~(\ref{dJ0}) and~(\ref{dJ3}) are
immediately derived.

\end{document}